\documentclass[twocolumn]{jpsj3}
\usepackage{txfonts}
\usepackage{graphicx}% Include figure files
\usepackage{dcolumn}% Align table columns on decimal point
\usepackage{bm}% bold math
\usepackage[usenames,dvipsnames]{xcolor}
\usepackage{ulem}
\usepackage{amsmath}
\usepackage{braket}
\usepackage{enumerate}
\usepackage{arydshln}
\usepackage{multirow}
\usepackage{booktabs}

\title{
Complete Multipole Basis Set for Single-Centered Electron Systems
}

\author{Hiroaki Kusunose$^{1}$, Rikuto Oiwa$^{1}$, and Satoru Hayami$^{2}$}
\inst{
$^1$Department of Physics, Meiji University, Kanagawa 214-8571, Japan \\
$^2$Department of Applied Physics, The University of Tokyo, Bunkyo, Tokyo 113-8656, Japan
}

\abst{
A whole series of expressions for four species of multipoles (electric, magnetic, magnetic toroidal, and electric toroidal) is provided as a complete basis set to describe arbitrary single-centered spinful electron systems.
A compact formula to calculate matrix elements of these multipoles is also derived.
A visualization method of an electronic state characterized in terms of multipoles is proposed.
The complete basis set is useful to narrow down a candidate order parameter of electron systems in phase transition, to describe a property of cross-correlated phenomena, to analyze spectra of x-ray scattering in magnetically ordered states, and so on.
We demonstrate a usage of the complete basis set by taking monopole and toroidal dipole orderings, and the mutual relationship among three distinct magnetic dipoles (orbital, spin angular momenta and anisotropic dipole) in a spin-orbit coupled system as prime examples.
}

\begin{document}
\maketitle

\section{Introduction}

A concept of multipole is widely used in various fields of physics such as classical electromagnetism~\cite{dubovik1975multipole,LandauLifshitz198001,nanz2016toroidal}, nuclear physics~\cite{Blin-Stoyle_RevModPhys.28.75,zel1958relation,Blatt1991}, solid-state physics~\cite{shiina1997magnetic,Kusunose_JPSJ.77.064710,kuramoto2008electronic,Santini_RevModPhys.81.807,kuramoto2009multipole,suzuki2018first,hayami2018microscopic,Hayami_PhysRevB.98.165110,Watanabe_PhysRevB.98.245129}, meta-materials~\cite{kaelberer2010toroidal,Savinov_PhysRevB.89.205112,papasimakis2016electromagnetic}, and so on.
In addition to well-known electric (E) and magnetic (M) multipoles in elementary electromagnetism, there exist magnetic toroidal (MT) and electric toroidal (ET) multipoles, which have common spatial parity to their time-reversal counterparts~\cite{dubovik1986axial,dubovik1990toroid,hayami2018microscopic,Hayami_PhysRevB.98.165110}.
Since these four species of multipoles are sufficient to describe arbitrary degrees of freedom of electromagnetic properties, they have been utilized to express multiple degrees of freedom of electrons in solids in accordance with symmetry.

In order to discuss electronic states microscopically in terms of multipoles, their quantum mechanical operator expressions are required.
In spinless systems, the operator expressions of E, M and MT multipoles have been obtained on the basis of the so-called multipole expansions of electromagnetic potentials~\cite{Spaldin_0953-8984-20-43-434203,kopaev2009toroidal,hayami2018microscopic}.
As for the remaining of ET multipole, which does not appear in the multipole expansion, its operator expression can be deduced from the time-reverting operation to M multipole~\cite{hayami2018microscopic}.
Moreover, such multipole expansions are straightforwardly extended to spinful systems by including the spin contribution to the electric current.

However, the multipoles introduced through the multipole expansions do not constitute a complete set, and quite a few multipoles are missing to satisfy the closure relation, especially for a spinful space.
Motivated by this circumstances, we provide a systematic definition of spinful multipoles of four species in this paper.
We derive a compact formula to calculate the matrix elements of a series of multipole operators with respect to total angular momentum basis or direct product of orbital and spin angular momentum bases.
Since 32 crystallographic point groups are subgroups of O(3) rotational group\cite{Inui1996}, operator expressions classified according to point-group symmetry are obtained merely by an appropriate linear combination of the expressions in the rotational group derived in this paper.

Once the whole expressions of multipole operators are obtained in a systematic way as in this paper, they are also useful to express cluster and bond extensions of multipoles~\cite{EdererPhysRevB.76.214404,Yanase_JPSJ.83.014703,hayami2016emergent,Suzuki_PhysRevB.95.094406,matsumoto2017symmetry,Suzuki_PhysRevB.99.174407,thole2018magnetoelectric,Hayami_PhysRevLett.122.147602}.
They are also utilized to describe the multipoles in momentum space~\cite{Hayami_PhysRevB.98.165110}, since the hopping integrals are essentially  single-centered quantity from their hopping origins.
The matrix elements of these augmented multipoles defined over a cluster are also obtained by the mapping between a sub-lattice and a molecular orbital basis of a cluster.

The organization of this paper is as follows.
In \S2, we first introduce the definition of spinless multipoles and their matrix elements in orbital angular momentum basis, and then we extend the discussion to spinful multipoles and their matrix elements.
We give explicit expressions of spinful multipoles up to rank $1$.
In \S3, we discuss the relation between multipoles defined by the multipole expansions and those in a complete set derived in the previous section.
It becomes clear which multipoles are missing in the multipole expansions.
In \S4, we propose two complementary ways of visualization of an electronic state, which are useful to grasp anisotropy of electronic states, and the mutual relationship among distinct multipole degrees of freedom.
In \S5, we demonstrate a practical usage of the complete basis set by taking the simplest system with total angular momenta, $J=1/2$ and $3/2$, in the $s$ and $p$ orbitals.
The final section summarizes the paper.
In three Appendices, we give detailed derivations of the matrix elements, and the relation between the multipoles in the expansions and those in a complete set.

\section{Complete Multipole Basis Set}

\subsection{For spinless systems}

First, let us summarize a complete multipole basis set for spinless systems.
We have already discussed in the literature~\cite{hayami2018microscopic} that four species of multipole operators can describe arbitrary electronic degrees of freedom in orbital states, which are characterized by the orbital angular momentum, $L$, and its component, $M$.
They are defined as
\begin{align}
&
\hat{Q}_{l,m}^{\rm (orb)}=O_{l,m},
\\&
\hat{M}_{l,m}^{\rm (orb)}=\frac{1}{2}\frac{2}{l+1}\left[(\bm{\nabla}O_{l,m})\cdot\hat{\bm{l}}+\hat{\bm{l}}\cdot(\bm{\nabla}O_{l,m})\right],
\\&
\hat{T}_{l,m}^{\rm (orb)}=\frac{1}{2}\frac{2}{(l+1)(l+2)}\left[(\bm{\nabla}O_{l,m})\cdot(\bm{r}\times\hat{\bm{l}})-(\hat{\bm{l}}\times\bm{r})\cdot(\bm{\nabla}O_{l,m})\right],
\\&
\hat{G}_{l,m}^{\rm (orb)}=\frac{1}{2}\frac{4i}{(l+1)^{2}(l+2)}\left[(\bm{\nabla}O_{l,m})\cdot\hat{\bm{l}}\,\hat{\bm{l}}^{2}-\hat{\bm{l}}^{2}\,\hat{\bm{l}}\cdot(\bm{\nabla}O_{l,m})\right],
\end{align}
where $\hat{\bm{l}}=-i(\bm{r}\times\bm{\nabla})$ is the dimensionless orbital angular momentum operator, and the prefactor $1/2$ is due to symmetrization of the operators.
Note that in these expressions $(\bm{\nabla}O_{l,m})$ should be understood that $\bm{\nabla}$ acts only on $O_{l,m}$, and
\begin{align}
O_{l,m}(\bm{r})=\sqrt{\frac{4\pi}{2l+1}}r^{l}Y_{l,m}(\hat{\bm{r}}),
\quad
\hat{\bm{r}}=\frac{\bm{r}}{r},
\end{align}
is proportional to the spherical harmonics $Y_{l,m}(\hat{\bm{r}})$ of the orbital angular momentum (rank of multipole), $l=0,1,2,\cdots$ and its $z$-component, $m=-l,-l+1,\cdots,l$.
We adopt the Racah normalization and the Condon-Shortley phase convention, \textit{i.e.}, $Y_{l,m}(\hat{\bm{r}})=(-1)^{m}\,Y_{l,-m}^{*}(\hat{\bm{r}})$.
Note that these multipole operators satisfy the relation,
\begin{align}
\biggl[\hat{X}_{l,m}^{\rm (orb)}\biggr]^{\dagger}=(-1)^{m}\,\hat{X}_{l,-m}^{\rm (orb)},
\quad
(X=Q,M,T,G).
\end{align}
Hereafter, we omit the hat symbol ($\hat{\,\,}$) for notational simplicity.

These multipoles have different parities with respect to time-reversal and spatial inversion operations.
Namely, the E multipole $Q_{l,m}$ and the MT multipole $T_{l,m}$ have definite spatial parity, $(-1)^{l}$ (called as polar or true tensor), while the M multipole $M_{l,m}$ and the ET multipole $G_{l,m}$ have parity $(-1)^{l+1}$ (called as axial or pseudo tensor).
Moreover, the ``electric'' and ``magnetic'' represent their time-reversal parities (even or odd).

By considering the fact that $X_{l,m}^{\rm (orb)}$ transforms like $Y_{l,m}$ with respect to the spatial rotation, the matrix element of the orbital angular momentum basis, $\ket{n_{i}L_{i}M_{i}}$ ($i=1,2$), can be decomposed as
\begin{multline}
\braket{n_{1}L_{1}M_{1}|X_{l,m}^{\rm (orb)}|n_{2}L_{2}M_{2}}
\\
=(-1)^{L_{1}+M_{1}}
\begin{pmatrix}
L_{1} & L_{2} & l \\ -M_{1} & M_{2} & m
\end{pmatrix}
\braket{n_{1}L_{1}||X_{l}^{\rm (orb)}||n_{2}L_{2}},
\label{wedecomp}
\end{multline}
where the parenthesis represents the Wigner's $3j$ symbol, and $\braket{n_{1}L_{1}||X_{l}^{\rm (orb)}||n_{2}L_{2}}$ is the so-called reduced matrix element.
Their explicit expressions are given below.
The additional index $n_{i}$ indicates quantum numbers other than $L_{i},M_{i}$ such as the principal quantum number.
This is a consequence of the Wigner-Eckart theorem.

The reduced matrix elements for $X=Q$ and $M$ are given by
\begin{align}
&
\braket{n_{1}L_{1}||Q_{l}^{\rm (orb)}||n_{2}L_{2}}
\cr&\quad
=(-1)^{L_{1}}\braket{r^{l}}_{12}\sqrt{(2L_{1}+1)(2L_{2}+1)}
\begin{pmatrix} L_{1} & L_{2} & l \\ 0 & 0 & 0 \end{pmatrix}
,
\label{rq}
\\&
\braket{n_{1}L_{1}||M_{l}^{\rm (orb)}||n_{2}L_{2}}
=(-1)^{L_{1}}\braket{r^{l-1}}_{12}\frac{2(2L_{2}+1)}{l+1}
\cr&\quad\quad\quad\times
\sqrt{l(2l+1)(2l-1)(2L_{1}+1)L_{2}(L_{2}+1)}
\cr&\quad\quad\quad\quad\quad\times
\begin{pmatrix} L_{1} & L_{2} & l-1 \\ 0 & 0 & 0 \end{pmatrix}
\begin{Bmatrix} l-1 & l & 1 \\ L_{2} & L_{2} & L_{1} \end{Bmatrix},
\label{rm}
\end{align}
respectively, where the curly bracket is the Wigner's $6j$ symbol, and
\begin{align}
\braket{r^{k}}_{12}=\int_{0}^{\infty}dr\,r^{k+2}R_{n_{1}L_{1}}(r)R_{n_{2}L_{2}}(r),
\end{align}
is the matrix element in the radial part.

The reduced matrix elements for $X=T$ and $G$ are proportional to those for $Q$ and $M$ as
\begin{align}
&
\braket{n_{1}L_{1}||T_{l}^{\rm (orb)}||n_{2}L_{2}}
=
i\,R_{l}(L_{1},L_{2})
\braket{n_{1}L_{1}||Q_{l}^{\rm (orb)}||n_{2}L_{2}},
\cr&
\braket{n_{1}L_{1}||G_{l}^{\rm (orb)}||n_{2}L_{2}}
=
i\,R_{l}(L_{1},L_{2})
\braket{n_{1}L_{1}||M_{l}^{\rm (orb)}||n_{2}L_{2}},
\label{relmt}
\end{align}
where the common proportional coefficient is given by
\begin{align}
R_{l}(L_{1},L_{2})=-\frac{L_{1}(L_{1}+1)-L_{2}(L_{2}+1)}{(l+1)(l+2)}.
\label{rl12}
\end{align}
From this expression, it is apparent that $T^{\rm (orb)}_{l,m}$ and $G^{\rm (orb)}_{l,m}$ are non-active for $L_{1}=L_{2}$.
Since the Wigner's symbols and the radial matrix elements are real, the matrix elements of $Q_{l,m}^{\rm (orb)}$ and $M_{l,m}^{\rm (orb)}$ are real, while those of $T_{l,m}^{\rm (orb)}$ and $G_{l,m}^{\rm (orb)}$ are pure imaginary.
The detailed derivation is given in Appendix~\ref{app1}.

\subsection{For spinful systems}

Next, we extend the complete multipole basis set to spinful (two-component spinor) systems.
The spinful space can be decomposed into charge and spin sectors; the $2\times2$ identity matrix $\sigma_{0}$ acts on charge sector, while the Pauli matrices (half of them are the dimensionless spin operators) $\bm{\sigma}=(\sigma_{x},\sigma_{y},\sigma_{z})$ act on spin sector, respectively.
Since $\sigma_{0}$ and $\bm{\sigma}$ are regarded as rank $0$ and $1$ tensors, it is natural to construct the spinful multipole operators by composing $\sigma_{0}$ and $\bm{\sigma}$ with $X_{l,m}^{\rm (orb)}$ in accordance with the addition rule of angular momentum.
The definition of the composed spinful multipole operators is given by
\begin{align}
X_{l,m}^{(s)}(k)&\equiv
i^{s+k}(-1)^{l+m}\sqrt{2l+1}
\cr&\quad\quad\times
\sum_{n=-s}^{s}
\begin{pmatrix} l+k & l & s \\ m-n & -m & n \end{pmatrix}
X_{l+k,m-n}^{\rm (orb)}\sigma_{s,n}.
\label{sfmpop}
\end{align}
Here, the indices $s=0$ and $k=0$ specify a multipole in charge sector with $\sigma_{0,0}=\sigma_{0}$, while $s=1$ and $k=-1,0,1$ specify that in spin sector where three spin components ($n=0,\pm1$) are defined as $\sigma_{1,0}=\sigma_{z}$ and  $\sigma_{1,\pm1}=\mp(\sigma_{x}\pm i\sigma_{y})/\sqrt{2}$.
Thanks to the phase factor $i^{s+k}$, the spinful multipole operator also satisfies
\begin{align}
\biggl[X_{l,m}^{(s)}(k)\biggr]^{\dagger}=(-1)^{m}\,X_{l,-m}^{(s)}(k).
\end{align}
It is easily confirmed that $X_{l,m}^{(0)}(0)=X_{l,m}^{\rm (orb)}\sigma_{0}$ in charge sector.

Since $\bm{\sigma}$ is the time-reversal odd axial vector, the time-reversal parity of $X_{l,m}^{(1)}(k)$ is opposite to that of $X_{l,m}^{\rm (orb)}$ for $k=0,\pm1$, and the spatial parity is opposite as well for $k=\pm1$ components.
The correspondence is summarized in Table~\ref{tbl1}.

\begin{table}[t!]
\caption{Correspondence between $X_{l}^{(1)}(k)$ and $X_{l+k}^{\rm (orb)}$ in spin sector.}
\vspace{3mm}
\label{tbl1}
\begin{center}
\renewcommand{\arraystretch}{1.4}
\begin{tabular}{ccc} \hline\hline
$X_{l}^{(1)}(k)$ & $k=0$ & $k=\pm1$ \\ \hline
$Q_{l}^{(1)}(k)$ & $T_{l}^{\rm (orb)}$ & $M_{l\pm1}^{\rm (orb)}$ \\
$M_{l}^{(1)}(k)$ & $G_{l}^{\rm (orb)}$ & $Q_{l\pm1}^{\rm (orb)}$ \\
$T_{l}^{(1)}(k)$ & $Q_{l}^{\rm (orb)}$ & $G_{l\pm1}^{\rm (orb)}$ \\
$G_{l}^{(1)}(k)$ & $M_{l}^{\rm (orb)}$ & $T_{l\pm1}^{\rm (orb)}$ \\
\hline\hline
\end{tabular}
\end{center}
\end{table}

In the presence of the spin-orbit coupling, it is natural to use the eigenstates  $\ket{JM;L}$ of the total angular momentum operator $\bm{j}=\bm{l}+\bm{\sigma}/2$ as a spinful basis.
They are explicitly given by
\begin{align}
&
\ket{JM;L}=(-1)^{J+M}\sqrt{2J+1}
\cr&\quad\quad\times
\sum_{\sigma}^{\pm1/2}
\begin{pmatrix} L & J & 1/2 \\ M-\sigma & -M & \sigma \end{pmatrix}
\ket{L,M-\sigma}\ket{\sigma}.
\end{align}
$J$ is positive half integer, and $M=-J,-J+1,\cdots,J$.
The orbital angular momentum $L$ in $\ket{JM;L}$ (non-negative integer) is omitted in what follows for notational simplicity.

In this basis, the matrix element of $X_{l,m}^{(s)}(k)$ is also decomposed as Eq.~(\ref{wedecomp}) by the Wigner-Eckart theorem, since $X_{l,m}^{(s)}(k)$ transforms like $Y_{l,m}$ under the spatial rotation:
\begin{multline}
\braket{n_{1}J_{1}M_{1}|X_{l,m}^{(s)}(k)|n_{2}J_{2}M_{2}}
\\
=(-1)^{J_{1}+M_{1}}
\begin{pmatrix}
J_{1} & J_{2} & l \\ -M_{1} & M_{2} & m
\end{pmatrix}
\braket{n_{1}J_{1}||X_{l}^{(s)}(k)||n_{2}J_{2}}.
\label{sfwe}
\end{multline}

The reduced matrix elements $\braket{n_{1}J_{1}||X_{l}^{(s)}(k)||n_{2}J_{2}}$ can be expressed in terms of $\braket{n_{1}L_{1}||X_{l}^{\rm (orb)}||n_{2}L_{2}}$ as defined in Eqs.~(\ref{rq}), (\ref{rm}), and (\ref{relmt}).
The detailed derivation is given in Appendix~\ref{app2}, and the results are given as follows.
In charge sector, it is given by
\begin{multline}
\braket{n_{1}J_{1}||X_{l}^{(0)}(0)||n_{2}J_{2}}
\\
=(-1)^{J_{1}+1/2+L_{2}+l}\sqrt{(2J_{1}+1)(2J_{2}+1)}\braket{n_{1}L_{1}||X_{l}^{\rm (orb)}||n_{2}L_{2}}
\\\times
\begin{Bmatrix} L_{1} & L_{2} & l \\ J_{2} & J_{1} & 1/2 \end{Bmatrix}.
\label{csex}
\end{multline}
In spin sector,
\begin{multline}
\braket{n_{1}J_{1}||X_{l}^{(1)}(k)||n_{2}J_{2}}
\\
=-i^{k+1}P_{l}^{(k)}(J_{1},J_{2};L_{1},L_{2})\braket{n_{1}L_{1}||X_{l+k}^{\rm (orb)}||n_{2}L_{2}},
\label{ssex}
\end{multline}
with
\begin{multline}
P_{l}^{(k)}(J_{1},J_{2};L_{1},L_{2})
=\sqrt{6(2l+1)(2J_{1}+1)(2J_{2}+1)}
\\\times
\begin{Bmatrix} L_{1} & J_{1} & 1/2 \\ L_{2} & J_{2} & 1/2 \\ l+k & l & 1 \end{Bmatrix},
\label{p12}
\end{multline}
where the curly bracket is the Wigner's $9j$ symbol.
Using these expressions, Eq.~(\ref{relmt}) and Table~\ref{tbl1}, we obtain explicit relations between $\braket{n_{1}J_{1}||X_{l}^{(1)}(k)||n_{2}J_{2}}$ and $\braket{n_{1}L_{1}||Q_{l+k}^{\rm (orb)}||n_{2}L_{2}}$ or $\braket{n_{1}L_{1}||M_{l+k}^{\rm (orb)}||n_{2}L_{2}}$ for $k=0$ and $k=\pm1$ as
\begin{align}
&
\braket{n_{1}J_{1}||Q_{l}^{(1)}(0)||n_{2}J_{2}}
=R_{l}\,P_{l}^{(0)}\,\braket{n_{1}L_{1}||Q_{l}^{\rm (orb)}||n_{2}L_{2}},
\cr&
\braket{n_{1}J_{1}||M_{l}^{(1)}(0)||n_{2}J_{2}}
=R_{l}\,P_{l}^{(0)}\,\braket{n_{1}L_{1}||M_{l}^{\rm (orb)}||n_{2}L_{2}},
\cr&
\braket{n_{1}J_{1}||T_{l}^{(1)}(0)||n_{2}J_{2}}
=-i\,P_{l}^{(0)}\,\braket{n_{1}L_{1}||Q_{l}^{\rm (orb)}||n_{2}L_{2}},
\cr&
\braket{n_{1}J_{1}||G_{l}^{(1)}(0)||n_{2}J_{2}}
=-i\,P_{l}^{(0)}\,\braket{n_{1}L_{1}||M_{l}^{\rm (orb)}||n_{2}L_{2}},
\label{rel1}
\end{align}
\begin{align}
&
\braket{n_{1}J_{1}||Q_{l}^{(1)}(\pm1)||n_{2}J_{2}}
=\pm P_{l}^{(\pm1)}\,\braket{n_{1}L_{1}||M_{l\pm1}^{\rm (orb)}||n_{2}L_{2}},
\cr&
\braket{n_{1}J_{1}||M_{l}^{(1)}(\pm1)||n_{2}J_{2}}
=\pm P_{l}^{(\pm1)}\,\braket{n_{1}L_{1}||Q_{l\pm1}^{\rm (orb)}||n_{2}L_{2}},
\cr&
\braket{n_{1}J_{1}||T_{l}^{(1)}(\pm1)||n_{2}J_{2}}
=\pm i\,R_{l\pm1}\,P_{l}^{(\pm1)}\,\braket{n_{1}L_{1}||M_{l\pm1}^{\rm (orb)}||n_{2}L_{2}},
\cr&
\braket{n_{1}J_{1}||G_{l}^{(1)}(\pm1)||n_{2}J_{2}}
=\pm i\,R_{l\pm1}\,P_{l}^{(\pm1)}\,\braket{n_{1}L_{1}||Q_{l\pm1}^{\rm (orb)}||n_{2}L_{2}},
\cr&
\label{rel2}
\end{align}
where we have omitted the common arguments of $P_{l}^{(k)}(J_{1},J_{2};L_{1},L_{2})$ and $R_{l}(L_{1},L_{2})$ for simplicity.
Eventually, the reduced matrix elements of $Q_{l+k}^{\rm (orb)}$ and $M_{l+k}^{\rm (orb)}$ are only required to calculate the matrix elements of multipole operators of any kind, which are given by Eqs.~(\ref{rq}) and (\ref{rm}) with use of Eqs.~(\ref{rl12}) and (\ref{p12}).
Since $R_{l}(L_{1},L_{2})$ vanishes for $L_{1}=L_{2}$, the active multipoles for $L_{1}=L_{2}$ are $T_{l,m}^{(1)}(0)$, $G_{l,m}^{(1)}(0)$, $Q_{l,m}^{(1)}(\pm1)$, and $M_{l,m}^{(1)}(\pm 1)$.

A set of spinful multipoles defined in the above, $\set{X_{l,m}^{(s)}(k)}$ ($X=Q,M,T,G$) constitutes a complete multipole basis set to describe arbitrary electronic degrees of freedom in single-centered electron systems.
Each multipole is specified by the indices $\alpha\equiv(s,k,l,m)$ where $s=0$ ($k=0$) or $s=1$ ($k=-1,0,+1$); $l=0,1,2,\cdots$; $m=-l,-l+1,\cdots,l$.
For practical purpose, it is convenient to normalize $X_{\alpha}$ so as to satisfy ${\rm Tr}(X_{\alpha}X_{\beta})=d\,\delta_{\alpha,\beta}$ with $d$ being the dimension of relevant Hilbert space for instance.

\subsection{For direct product of orbital and spin systems}

In the case of negligible spin-orbit coupling, it is convenient to use the electronic basis of the direct product, $\ket{LM\sigma}=\ket{LM}\otimes\ket{1/2\sigma}$.
Accordingly, the multipole operators are also defined as a direct product of $X_{l,m}^{\rm (orb)}$ and $\sigma_{1,n}$:
\begin{align}
X_{l,m}^{\otimes}(n)\equiv X_{l,m}^{\rm (orb)}\sigma_{1,n},
\quad
(n=0,\pm1),
\end{align}
where $X_{l,m}^{\otimes}(n)$ and $X_{l,m}^{\rm (orb)}$ share the same spatial parity with the opposite time-reversal property.
Namely, the correspondence between them is given by the column $k=0$ in Table~\ref{tbl1}.
The matrix elements are given by
\begin{align}
&
\braket{n_{1}L_{1}M_{1}\sigma_{1}|X_{l,m}^{\otimes}(n)|n_{2}L_{2}M_{2}\sigma_{2}}
\cr&\quad
=
\braket{n_{1}L_{1}M_{1}|X_{l,m}^{\rm (orb)}|n_{2}L_{2}M_{2}}
\braket{\sigma_{1}|\sigma_{1,n}|\sigma_{2}},
\\&
\braket{\sigma_{1}|\sigma_{1,n}|\sigma_{2}}=
(-1)^{\sigma_{1}-1/2}\sqrt{6}
\begin{pmatrix} 1/2 & 1/2 & 1 \\ -\sigma_{1} & \sigma_{2} & n \end{pmatrix},
\end{align}
where $\braket{n_{1}L_{1}M_{1}|X_{l,m}^{\rm (orb)}|n_{2}L_{2}M_{2}}$ is given by Eq.~(\ref{wedecomp}).

\subsection{Explicit Expressions up to Rank $1$}\label{expmul}

Here, we give explicit expressions of active multipoles up to rank $1$ by using the definition, Eq.~(\ref{sfmpop}).

\begin{itemize}
\item Monopole : $X^{(s)}(k)\equiv X_{0,0}^{(s)}(k)$.
\begin{align}
&
Q^{(0)}(0)=\sigma_{0},
\quad
Q^{(1)}(1)=\frac{1}{\sqrt{3}}(\bm{l}\cdot\bm{\sigma}),
\\&
M^{(1)}(1)=\frac{1}{\sqrt{3}}(\bm{r}\cdot\bm{\sigma}),
\\&
G^{(1)}(1)=\frac{1}{\sqrt{3}}(\bm{t}\cdot\bm{\sigma}),
\quad
\bm{t}=\frac{1}{6}[(\bm{r}\times\bm{l})-(\bm{l}\times\bm{r})].
\end{align}
\item Dipole : $\displaystyle \bm{X}^{(s)}(k)\equiv\sum_{m=-1}^{1}X_{1,m}^{(s)}(k)\,\bm{e}_{1,m}^{*}$.
\begin{align}
&
\bm{Q}^{(0)}(0)=\bm{r}\sigma_{0},
\quad
\bm{Q}^{(1)}(0)=\frac{1}{\sqrt{2}}(\bm{\sigma}\times\bm{t}),
\cr&
[\bm{Q}^{(1)}(1)]_{z}
=\frac{2}{3\sqrt{10}}\biggl[3(lr)_{zx}\sigma_{x}+3(lr)_{yz}\sigma_{y}
\cr&\quad\quad
+2[2(lr)_{zz}-(lr)_{xx}-(lr)_{yy}]\sigma_{z}\biggr]
\quad(+{\rm cyclic}),
\\&
\bm{M}^{(0)}(0)=\bm{l}\sigma_{0},
\quad
\bm{M}^{(1)}(-1)=\bm{\sigma},
\cr&
\bm{M}^{(1)}(1)=\frac{3}{\sqrt{10}}\left[(\bm{r}\cdot\bm{\sigma})\bm{r}-\frac{\bm{r}^{2}}{3}\bm{\sigma}\right],
\\&
\bm{T}^{(0)}(0)=\bm{t}\sigma_{0},
\quad
\bm{T}^{(1)}(0)=\frac{1}{\sqrt{2}}(\bm{\sigma}\times\bm{r}),
\cr&
[\bm{T}^{(1)}(1)]_{z}
=\frac{1}{2\sqrt{10}}\biggl[3(lt)_{zx}\sigma_{x}+3(lt)_{yz}\sigma_{y}
\cr&\quad\quad
+2[2(lt)_{zz}-(lt)_{xx}-(lt)_{yy}]\sigma_{z}\biggr]\quad(+{\rm cyclic}),
\\&
\bm{G}^{(1)}(0)=\frac{1}{\sqrt{2}}(\bm{\sigma}\times\bm{l}),
\cr&
[\bm{G}^{(1)}(1)]_{z}
=\frac{1}{2\sqrt{10}}\biggl[3(tr)_{zx}\sigma_{x}+3(tr)_{yz}\sigma_{y}
\cr&\quad\quad
+2[2(tr)_{zz}-(tr)_{xx}-(tr)_{yy}]\sigma_{z}\biggr]\quad(+{\rm cyclic}).
\end{align}
\end{itemize}
Here, $\bm{e}_{1,0}=\bm{e}_{z}$  and $\bm{e}_{1,\pm1}=\mp(\bm{e}_{x}\pm i\bm{e}_{y})/\sqrt{2}$, and $(AB)_{ij}=(A_{i}B_{j}+A_{j}B_{i}+B_{i}A_{j}+B_{j}A_{i})/4$, and $\bm{e}_{i}$ ($i=x,y,z$) are unit vectors in the Euclidean coordinate.

The E monopole $Q^{(1)}(1)$ is nothing but the atomic spin-orbit coupling, and the M monopole $M^{(1)}(1)$ is the atomic limit of the so-called magnetic flux.
The ET monopole $G^{(1)}(1)$ is an essential ingredient for chiral and strong gyrotropic point groups.

$\bm{M}^{(1)}(1)$ represents the anisotropic magnetic dipole operator, which is independent of the ordinary orbital and spin angular momentum operators, $\bm{M}^{(0)}(0)=\bm{l}$ and $\bm{M}^{(1)}(-1)=\bm{\sigma}$.
This type of the anisotropic magnetic dipole has been known to appear in the context of the x-ray magneto-circular dichroism (XMCD), referred as $\bm{T}$-vector in literatures~\cite{Carra_PhysRevLett.70.694,stohr1995x,Stohr_PhysRevLett.75.3748,crocombette1996importance}.

The MT dipole $\bm{T}^{(s)}(k)$ is relevant to the time-reversal odd axial tensor such as the linear magneto-electric effect~\cite{Spaldin_0953-8984-20-43-434203}.
The ET vector $\bm{G}^{(s)}(k)$ appears in the off-diagonal elements of the time-reversal even polar tensor such as Seebeck effect~\cite{Hayami_PhysRevB.98.165110}.

\section{Relation to Multipoles Defined by Potential Expansion}

Multipoles usually appear in the multipole expansions of the scalar and vector potentials~\cite{Schwartz_PhysRev.97.380,dubovik1990toroid,Blatt1991,Kusunose_JPSJ.77.064710}.
In other words, the multipoles $X_{l,m}$ are defined through the expansions.
In the Coulomb gauge $\bm{\nabla}\cdot\bm{A}=0$, the expansions in unit of $-e=-\mu_{\rm B}=1$ are given by
\begin{align}
&
\phi(\bm{r})=\sum_{lm}\sqrt{\frac{4\pi}{2l+1}}\braket{Q_{l,m}}\frac{Y_{l,m}^{*}(\hat{\bm{r}})}{r^{l+1}},
\label{pphi}
\\&
\bm{A}(\bm{r})=\sum_{lm}\left[i\sqrt{\frac{4\pi}{(2l+1)l}}\braket{M_{l,m}}\frac{\bm{Y}^{l*}_{l,m}(\hat{\bm{r}})}{r^{l+1}}
\right.\cr&\quad\quad\quad\quad\quad\left.
-\sqrt{4\pi(l+1)}\braket{T_{l,m}}\frac{\bm{Y}^{l+1*}_{l,m}(\hat{\bm{r}})}{r^{l+2}}\right],
\label{pa}
\end{align}
where $\bm{Y}_{l,m}^{l+k}(\hat{\bm{r}})$ ($k=0,\pm1$) is the vector spherical harmonics, whose definition is given by Eq.~(\ref{vsh}), and $\braket{X_{l,m}}$ indicates an appropriate thermal average of the multipole operator $X_{l,m}$.
Note that which component of multipoles appears in the expansions depends on the gauge fixing condition.
The ET multipole $G_{l,m}$ does not appear in the expansion.
In order to define the ET multipole operator in a systematic way, the operator $R_{T}=\bm{t}_{l}\cdot\bm{\nabla}$, which reverts the time-reversal parity with keeping the spatial parity, is used as $G_{l,m}\equiv R_{T}Q_{l,m}$, where $\bm{t}_{l}$ is the elementary MT dipole~\cite{hayami2018microscopic}.

The multipole operators $X_{l,m}$ in the context of the potential expansions \textit{do not} constitute a complete basis set, and quite a few spinful multipoles are missing.
In fact, the multipole $X_{l,m}$ hereby can be expanded by the spinful multipole $X_{l,m}^{(s)}(k)$ as follows (see Appendix~\ref{app3} in detail):
\begin{align}
Q_{l,m}&=Q_{l,m}^{\rm (orb)}\sigma_{0}=Q_{l,m}^{(0)}(0),
\\
M_{l,m}&=M_{l,m}^{\rm (orb)}\sigma_{0}+(\bm{\nabla}O_{l,m})\cdot\bm{\sigma}
\cr&
=M_{l,m}^{(0)}(0)+\sqrt{l(2l-1)}M_{l,m}^{(1)}(-1),
\label{mpm}
\\
T_{l,m}&=T_{l,m}^{\rm (orb)}\sigma_{0}
\cr&\quad
+\frac{1}{2}\frac{1}{l+1}\left[
(\bm{\nabla}O_{l,m})\cdot(\bm{r}\times\bm{\sigma})
+(\bm{r}\times\bm{\sigma})\cdot(\bm{\nabla}O_{l,m})
\right]
\cr&
=T_{l,m}^{(0)}(0)-\sqrt{\frac{l}{l+1}}T_{l,m}^{(1)}(0),
\label{mpt}
\\
G_{l,m}&=G_{l,m}^{\rm (orb)}\sigma_{0}+\frac{1}{2}\sum_{ij}^{x,y,z}\left[(\nabla_{i}\nabla_{j}O_{l,m})g_{ij}+g_{ij}^{\dagger}(\nabla_{i}\nabla_{j}O_{l,m})\right]
\cr&
=G_{l,m}^{(0)}(0)+\frac{l\sqrt{l(2l-1)}}{l+2}G_{l,m}^{(1)}(-1)-\sqrt{\frac{l}{l+1}}G_{l,m}^{(1)}(0),
\label{mpg}
\quad\quad
\end{align}
where
\begin{align}
g_{ij}=\frac{2(\bm{r}\times\bm{l})_{i}\sigma_{j}}{(l+1)(l+2)}
+\frac{2(\bm{r}\times\bm{\sigma})_{i}l_{j}}{(l+1)^{2}}.
\label{gfac}
\end{align}
Although there exists the term proportional to $(\bm{r}\times\bm{\sigma})_{i}\sigma_{j}$ in the definition of $G_{l,m}$, it is shown to vanish identically.

It should be emphasized that the expressions of $X_{l,m}$ were obtained by assuming that the sources of $\phi(\bm{r})$ and $\bm{A}(\bm{r})$ are ordinary density of electron charge $\rho(\bm{r})$ and orbital and spin contributions to the electric current $\bm{j}(\bm{r})=\bm{j}_{\rm orb}(\bm{r})+\bm{j}_{\rm spin}(\bm{r})$.
However, other multipoles which do not appear in the expansions are also able to be an order parameter through many-body interactions.
Once such an order parameter appears, it induces other multipoles belonging to the same irreducible representation of it.
Eventually, other multipoles such as $Q_{l,m}^{(1)}(k)$ and $G_{l,m}^{(1)}(+1)$ must contribute to $\phi(\bm{r})$, and $M_{l,m}^{(1)}(+1)$, and $T_{l,m}^{(1)}(\pm 1)$ to $\bm{A}(\bm{r})$ in this sense.

\section{Visualization of Electronic States}

When a thermal average of the specific multipole operator becomes finite, the corresponding anisotropy appears.
Thus, it is useful to visualize the anisotropy that characterizes an electronic state in the presence of multipoles~\cite{Kusunose_JPSJ.77.064710}.
We show two complementary ways to visualize the anisotropy of a system having eigenstates $\ket{\psi_{\gamma}}$ with the energy $E_{\gamma}$: The anisotropy is expressed (i) by a distribution of electric charge and current, and (ii) by a distribution of elementary multipole charges.

\subsection{Based on charge and angular-momentum distributions}

A thermal average of an operator $A$ at inverse temperature $\beta$ is given by
\begin{align}
\braket{A}=\frac{1}{Z}\sum_{\gamma}e^{-\beta E_{\gamma}}\braket{\psi_{\gamma}|A|\psi_{\gamma}},
\quad
Z=\sum_{\gamma}e^{-\beta E_{\gamma}}.
\end{align}
By using the total angular-momentum basis ($i=nJM;L$) as $\ket{\psi_{\gamma}}=\sum_{i}U_{i\gamma}\ket{i}$, the expectation value of $A$ with respect to the $\gamma$ eigenstate is expressed as
\begin{align}
&
\braket{\psi_{\gamma}|A|\psi_{\gamma}}=\sum_{i_{2}i_{3}}U_{i_{3}\gamma}^{*}U_{i_{2}\gamma}^{}\braket{i_{3}|A|i_{2}}
\cr&\quad
=\sum_{i_{2}i_{3}}U_{i_{3}\gamma}^{*}U_{i_{2}\gamma}^{}\sum_{\sigma}\int d\hat{\bm{r}}\,
\sum_{i_{1}}
\psi_{i_{3}}^{*}(\hat{\bm{r}},\sigma)\psi_{i_{1}}^{}(\hat{\bm{r}},\sigma)\braket{i_{1}|A|i_{2}}
\cr&\quad
=\int \frac{d\hat{\bm{r}}}{4\pi}\,\,\overline{A}_{\gamma}(\hat{\bm{r}}),
\end{align}
where we have inserted the completeness relation, $1=\sum_{\sigma}\int d\hat{\bm{r}}\ket{\hat{\bm{r}}\sigma}\bra{\hat{\bm{r}}\sigma}\sum_{i_{1}}\ket{i_{1}}\bra{i_{1}}$, and 
\begin{align}
\psi_{i}(\hat{\bm{r}},\sigma)&=\braket{\hat{\bm{r}}\sigma|i}
=(-1)^{J+M}\sqrt{2J+1}
\cr&\quad\times
\begin{pmatrix} L & J & 1/2 \\ M-\sigma & -M & \sigma \end{pmatrix}
Y_{L,M-\sigma}(\hat{\bm{r}}).
\quad
\end{align}
Then, the angular distribution is given by
\begin{align}
&
\overline{A}_{\gamma}(\hat{\bm{r}})=\sum_{i_{1}i_{2}i_{3}}{\rm Re}\biggl[
U_{i_{3}\gamma}^{*}U_{i_{2}\gamma}^{}
\braket{i_{1}|A|i_{2}}P_{i_{3}i_{1}}(\hat{\bm{r}})
\biggr],
\cr&
P_{i_{3}i_{1}}(\hat{\bm{r}})=4\pi\sum_{\sigma}\psi_{i_{3}}^{*}(\hat{\bm{r}},\sigma)\psi_{i_{1}}^{}(\hat{\bm{r}},\sigma)
\cr&\quad
=(-1)^{J_{3}+M_{3}+J_{1}+M_{1}}4\pi\sqrt{(2J_{3}+1)(2J_{1}+1)}
\cr&\quad\times
\sum_{\sigma}
\begin{pmatrix} L_{3} & J_{3} & 1/2 \\ M_{3}-\sigma & -M_{3} & \sigma \end{pmatrix}
\begin{pmatrix} L_{1} & J_{1} & 1/2 \\ M_{1}-\sigma & -M_{1} & \sigma \end{pmatrix}
\cr&\quad\quad\quad\times
Y_{L_{3},M_{3}-\sigma}^{*}(\hat{\bm{r}})Y_{L_{1},M_{1}-\sigma}^{}(\hat{\bm{r}}).
\label{angledist2}
\end{align}
The angular distribution of the thermal average $\braket{A}$ is thus given by
\begin{align}
\overline{A}(\hat{\bm{r}})=\frac{1}{Z}\sum_{\gamma}e^{-\beta E_{\gamma}}\overline{A}_{\gamma}(\hat{\bm{r}}).
\end{align}
In the case of spinless systems, we replace $i$ and $P_{i_{3}i_{1}}(\hat{\bm{r}})$ in Eq.~(\ref{angledist2}) with $(i=nLM)$ and $P_{i_{3}i_{1}}(\hat{\bm{r}})=4\pi Y_{L_{3},M_{3}}^{*}(\hat{\bm{r}})Y_{L_{1},M_{1}}^{}(\hat{\bm{r}})$.

In order to visualize an electronic state, the angular distribution of specific multipole operator, such as the E charge $\rho=Q^{(0)}(0)$, and the M dipoles $\bm{l}=\bm{M}^{(0)}(0)$, $\bm{\sigma}=\bm{M}^{(1)}(-1)$, is useful.
These operators are given in \S\ref{expmul}.
For instance, $\overline{\rho}(\hat{\bm{r}})$ is used to express the shape of the wavefunction, and its magnetic property is displayed by the M dipole distributions.

\subsection{Based on multipole charge densities}

The thermal average of the multipoles is related with the polarization or magnetization density $\bm{X}(\bm{r})$ as
\begin{align}
\braket{X_{l,m}}
=\int d\bm{r}\,\bm{X}(\bm{r})\cdot(\bm{\nabla}O_{l,m}^{*})
=\int d\bm{r}\,O_{l,m}^{*}(\bm{r})\rho_{X}(\bm{r}),
\end{align}
where we have introduced the corresponding multipole ``charge'' density as $\rho_{X}(\bm{r})\equiv -\bm{\nabla}\cdot\bm{X}(\bm{r})$.
Here, $X_{l,m}$ is either $X_{l,m}^{(s)}(k)$, $X_{l,m}^{\otimes}(n)$ or $X_{l,m}$.
This is utilized to visualize an electronic state as follows.
To this end, we decompose $\rho_{X}(\bm{r})$ into the radial part $\rho_{X}(r)$ and angular part $\overline{\rho}_{X}(\hat{\bm{r}})$ as $\rho_{X}(\bm{r})=\rho_{X}(r)\overline{\rho}_{X}(\hat{\bm{r}})/4\pi$.
The latter can be extracted by the completeness relation of the spherical harmonics as
\begin{align}
\overline{\rho}_{X}(\hat{\bm{r}})=\sum_{lm}(2l+1)\frac{\braket{X_{l,m}}}{\braket{r^{l}}}\frac{O_{l,m}(\hat{\bm{r}})}{r^{l}},
\end{align}
where we have introduced the radial average $\braket{r^{l}}\equiv\int_{0}^{\infty} dr\,r^{l+2}\rho_{X}(r)$, which is roughly given by the average of $\braket{r^{l}}_{12}$ over relevant orbitals, $L_{1}$ and $L_{2}$.
For example, $\overline{\rho}_{Q}(\hat{\bm{r}})$ is used to express the shape of the wavefunction, and its M, MT, and ET charge distributions, \textit{i.e.}, $\overline{\rho}_{M}(\hat{\bm{r}})$, $\overline{\rho}_{T}(\hat{\bm{r}})$,  and $\overline{\rho}_{G}(\hat{\bm{r}})$ are plotted as a colormap on the surface of the shape.

\section{Examples}

\begin{figure}[h!]
\centering
\includegraphics[width=8.5cm]{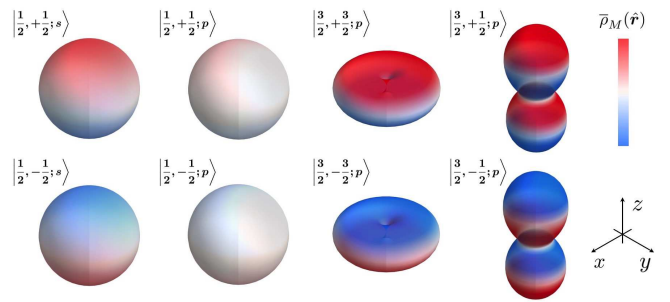}
\caption{Charge and magnetic charge densities of the wavefunctions $\ket{J,M;L}$.
In calculating $\overline{\rho}_{X}(\hat{\bm{r}})$, the multipoles in the potential expansions are used.
}
\label{basiswf}
\end{figure}

In order to demonstrate the complete multipole basis set, we consider $J=1/2$ and $J=3/2$ systems with $L=0$ ($s$) and $1$ ($p$) orbitals as an example.
The eigenstates of the total angular momentum $\ket{J,M;L}$ are given in terms of the linear combinations of the direct product of the orbital $\ket{L,M\pm\sigma}$ and spin $\ket{\sigma}$ states as
\begin{align}
&
\Ket{\frac{1}{2},+\frac{1}{2};s}=\ket{0,0}\ket{\uparrow},
\cr&
\Ket{\frac{1}{2},-\frac{1}{2};s}=\ket{0,0}\ket{\downarrow},
\\&
\Ket{\frac{1}{2},+\frac{1}{2};p}=\sqrt{\frac{2}{3}}\ket{1,+1}\ket{\downarrow}-\sqrt{\frac{1}{3}}\ket{1,0}\ket{\downarrow},
\cr&
\Ket{\frac{1}{2},-\frac{1}{2};p}=-\sqrt{\frac{2}{3}}\ket{1,-1}\ket{\uparrow}+\sqrt{\frac{1}{3}}\ket{1,0}\ket{\uparrow},
\\&
\Ket{\frac{3}{2},+\frac{3}{2};p}=\ket{1,+1}\ket{\uparrow},
\cr&
\Ket{\frac{3}{2},+\frac{1}{2};p}=\sqrt{\frac{1}{3}}\ket{1,+1}\ket{\downarrow}+\sqrt{\frac{2}{3}}\ket{1,0}\ket{\uparrow},
\cr&
\Ket{\frac{3}{2},-\frac{1}{2};p}=\sqrt{\frac{1}{3}}\ket{1,-1}\ket{\uparrow}+\sqrt{\frac{2}{3}}\ket{1,0}\ket{\downarrow},
\cr&
\Ket{\frac{3}{2},-\frac{3}{2};p}=\ket{1,-1}\ket{\downarrow}.
\end{align}
The wavefunctions of the basis $\ket{J,M;L}$ are visualized in Fig.~\ref{basiswf}, where $\overline{\rho}_{Q}(\hat{\bm{r}})$ and $\overline{\rho}_{M}(\hat{\bm{r}})$ are used for the shape and colormap, respectively.

\begin{table}[t!]
\caption{Active multipoles in $J=1/2(s)$, $1/2(p)$, and $3/2(p)$.
The upper (lower) off-diagonal represents electric (magnetic) multipoles.
We use the abbreviations, $\rho=Q^{(0)}(0)$, $G_{\sigma}=G^{(1)}(1)$, $M_{\sigma}=M^{(1)}(1)$, $\bm{r}=\bm{Q}^{(0)}(0)$, $\bm{Q}_{\sigma}=\bm{Q}^{(1)}(0)$, $\bm{l}=\bm{M}^{(0)}(0)$, $\bm{\sigma}=\bm{M}^{(1)}(-1)$, $\bm{M}_{\rm a}=\bm{M}^{(1)}(1)$, $\bm{t}=\bm{T}^{(0)}(0)$, $\bm{T}_{\sigma}=\bm{T}^{(1)}(0)$, and $\bm{G}_{\sigma}=\bm{G}^{(1)}(0)$.
$X_{2}^{(s)}(k)$ and $X_{3}^{(s)}(k)$ represent quadrupole and octupole, respectively.
}
\vspace{3mm}
\label{tbl2a}
\begin{center}
\renewcommand{\arraystretch}{1.5}
\begin{tabular}{l|l|l;{3pt/3pt}l}
$J=$ & $1/2$ $(s)$ & $1/2$ $(p)$ & $3/2$ $(p)$ \\ \hline
 $1/2$ $(s)$ & $\rho$ & $G_{\sigma}$ & $\bm{Q}_{\sigma}$ \\
& $\bm{\sigma}$ & $\bm{r}$ & $G_{2}^{(1)}(-1)$ \\ \hline
 $1/2$ $(p)$ & $M_{\sigma}$ & $\rho$ & $\bm{G}_{\sigma}$ \\
& $\bm{t}$ & $\bm{l}$ & $Q_{2}^{(1)}(-1)$ \\ \hdashline[3pt/3pt]
$3/2$ $(p)$ & $M_{2}^{(1)}(-1)$ & $\bm{M}_{\rm a}$ & $Q^{(1)}(1)$, $Q_{2}^{(0)}(0)$ \\
& $\bm{T}_{\sigma}$ & $T_{2}^{(1)}(0)$ & $\bm{\sigma}$, $M_{3}^{(1)}(-1)$ \\
\end{tabular}
\end{center}
\end{table}

The active multipoles in this Hilbert space are summarized in Table~\ref{tbl2a}, and their matrix elements are given in Ref.~\citen{comment_sm_mat}.
Among these multipoles, we focus on (i) the M and ET monopoles, $M_{\sigma}\equiv M^{(1)}(1)$ and $G_{\sigma}\equiv G^{(1)}(1)$, (ii) the MT and ET dipoles, $\bm{T}_{\sigma}\equiv\bm{T}^{(1)}(0)$ and $\bm{G}_{\sigma}\equiv \bm{G}^{(1)}(0)$, and (iii) the anisotropic M dipole, $\bm{M}_{\rm a}\equiv \bm{M}^{(1)}(1)$.

\subsection{The M and ET monopoles}

First, we consider the ground states of the following Hamiltonians,
\begin{align}
H_{\rm M0}=-M_{\sigma},
\quad
H_{\rm ET0}=-G_{\sigma}.
\end{align}
The ground states are doubly degenerate, and they are given by
\begin{align}
M_{\sigma}:&\,\,\,
\ket{\pm}=\frac{1}{\sqrt{2}}\left[\Ket{\frac{1}{2},\pm\frac{1}{2};s}-\Ket{\frac{1}{2}\pm\frac{1}{2};p}\right],
\\
G_{\sigma}:&\,\,\,
\ket{\pm}=\frac{1}{\sqrt{2}}\left[\Ket{\frac{1}{2},\pm\frac{1}{2};s}+ i\Ket{\frac{1}{2}\pm\frac{1}{2};p}\right],
\end{align}
respectively.

\begin{table}[t!]
\caption{Angular distributions for the M- and ET-monopole, and the MT- and ET-dipole eigenstates.
$\bm{X}$ is the essential vector that characterizes the wavefunction.}
\vspace{3mm}
\label{tbl2}
\begin{center}
\renewcommand{\arraystretch}{1.5}
\begin{tabular}{cccc} \hline\hline
$X_{l,m}^{(s)}(k)$ & $\overline{\rho}(\hat{\bm{r}})$ & $\bm{X}$ & $\overline{X}(\hat{\bm{r}})$ \\ \hline
$M_{\sigma}$ & $1$ & $\bm{\sigma}$ & $\hat{\bm{r}}$ \\
$G_{\sigma}$ & $1$ & $\bm{G}_{\sigma}$ & $\displaystyle-\frac{1}{\sqrt{2}}\hat{\bm{r}}$ \\ \hline
$T_{\sigma}^{z}$ & $\displaystyle\frac{1}{4}(5-3\hat{z}^{2})$ & $\bm{\sigma}$ & $\displaystyle\frac{\sqrt{6}}{2}(\hat{y},-\hat{x},0)$ \\
$G_{\sigma}^{z}$ & $\displaystyle\frac{1}{4}(3\hat{z}^{2}-1)$ & $\bm{Q}_{\sigma}$ & $\displaystyle\frac{-1}{12\sqrt{2}}(\hat{x}-\sqrt{2}\hat{y},\hat{y}+\sqrt{2}\hat{x},0)$ \\
\hline\hline
\end{tabular}
\end{center}
\end{table}

\begin{figure}[h!]
\centering
\includegraphics[width=7.5cm]{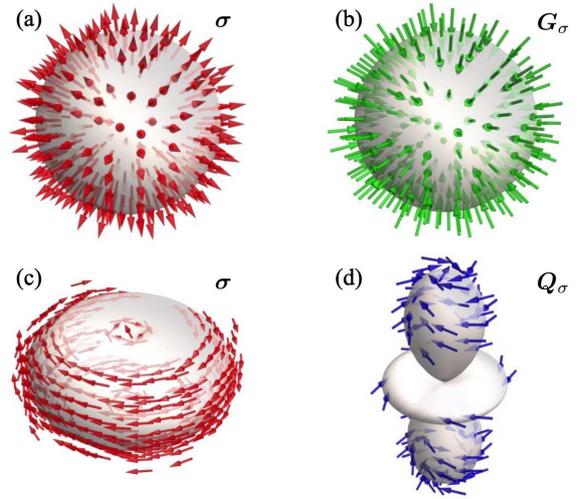}
\caption{Characteristic angular distributions for the ground states of (a) $M_{\sigma}$, (b) $G_{\sigma}$, (c) $T_{\sigma}^{z}$, and (d) $G_{\sigma}^{z}$.
The shape and arrows represent the charge distribution $\overline{\rho}(\hat{\bm{r}})$, and the angular distribution of $\bm{\sigma}$, $\bm{G}_{\sigma}$ or $\bm{Q}_{\sigma}$, respectively.
}
\label{angledist}
\end{figure}

For these ground states, angular distributions are summarized in the upper rows in Table~\ref{tbl2}.
Both in the M- and ET-monopole ground states, the angular distribution of $\rho$ is isotropic due to the nature of $J=1/2$ wavefunctions in Fig.~\ref{basiswf}.
In the M-monopole $M_{\sigma}$ ground state, the monopole flux appears in $\overline{\bm{\sigma}}(\hat{\bm{r}})$ as shown in Fig.~\ref{angledist}(a).
On the other hand, in the ET-monopole $G_{\sigma}$ ground state, there are no characteristic angular distributions in ordinary physical quantities such as the E dipole $\bm{r}$, and the M dipoles $\bm{l}$ and $\bm{\sigma}$.
Since the ET monopole is pseudoscalar, there must exist a monopole flux.
Indeed, it appears in the angular distribution of $\bm{G}_{\sigma}=(\bm{\sigma}\times\bm{l})/\sqrt{2}$ as shown in Fig.~\ref{angledist}(b).
Although it is difficult to observe directly $\bm{G}_{\sigma}$ itself, the ET monopole flux results in hedgehog-type spin polarization in momentum space in periodic systems, since $\bm{G}_{\sigma} \cdot \bm{r} \sim (\bm{\sigma}\times \bm{l}) \cdot \bm{r} =(\bm{l}\times \bm{r}) \cdot \bm{\sigma} \sim \bm{t} \cdot \bm{\sigma}$ has the same symmetry as $\bm{k} \cdot \bm{\sigma}$ where $\bm{k}$ is the wave vector.
Thus, an observation of the hedgehog spin textures in momentum space is helpful to identify the emergence of the ET monopole.

\subsection{The MT and ET dipoles}

Next, we consider the ground states of the following Hamiltonians,
\begin{align}
H_{\rm MT1}=-T_{\sigma}^{z},
\quad
H_{\rm ET1}=-G_{\sigma}^{z}.
\end{align}
The ground states are doubly degenerate, and they are given by
\begin{align}
T_{\sigma}^{z}:&\,\,\,
\ket{\pm}=\frac{1}{\sqrt{2}}\left[\Ket{\frac{1}{2},\pm\frac{1}{2};s}
\pm\frac{i}{\sqrt{3}}\Ket{\frac{1}{2}\pm\frac{1}{2};p}
\right.\cr&\quad\quad\quad\quad\left.
+\frac{i}{\sqrt{6}}\Ket{\frac{3}{2}\pm\frac{1}{2};p}\right],
\\
G_{\sigma}^{z}:&\,\,\,
\ket{\pm}=\frac{1}{\sqrt{2}}\left[\Ket{\frac{1}{2},\pm\frac{1}{2};p}+ i\Ket{\frac{3}{2}\pm\frac{1}{2};p}\right],
\end{align}
respectively.

In the MT-dipole $T_{\sigma}^{z}$ ground state, the vortex-like angular distribution of $\bm{\sigma}$ arises perpendicular to $\bm{T}_{\sigma}$ as shown in Fig.~\ref{angledist}(c).
It is characterized by the vorticity, $\int d\hat{\bm{r}}\,\hat{\bm{r}}\times\overline{\bm{\sigma}}(\hat{\bm{r}})\propto -\bm{e}_{z}$.
The charge distribution becomes anisotropic since the ground state is the superposition of anisotropic wavefunctions in Fig.~\ref{basiswf}.
The angular dependences are given in the lower rows in Table~\ref{tbl2}.
Similarly, the wavefunction of the ET-dipole $G_{\sigma}^{z}$ ground state becomes anisotropic as shown in Fig.~\ref{angledist}(d).
However, there are no indications of the toroidal nature in ordinary physical quantities.
Interestingly, it appears in less ordinary E-dipole involving spin degrees of freedom, namely, $\bm{Q}_{\sigma}=(\bm{\sigma}\times\bm{t})/\sqrt{2}$.
By looking at this quantity, we realize the toroidal nature in this ground state as shown in Fig.~\ref{angledist}(d).

In this way, the spinful multipoles are required to characterize the multipole ordered states in general.

\subsection{The anisotropic M dipole}

The anisotropic M dipole is often significant in analysis of XMCD spectra~\cite{Carra_PhysRevLett.70.694,stohr1995x,Stohr_PhysRevLett.75.3748,crocombette1996importance,yamasaki2019augmented}.
In such a context, the relation among orbital wavefunctions in a certain principal axis and three distinct M dipoles, $\bm{l}$, $\bm{\sigma}$, and $\bm{M}_{\rm a }\equiv \bm{M}^{(1)}(1)$ provides useful information.

To elucidate the relation between them, let us consider a simplified situation in a magnetically ordered state, which is described by the following Hamiltonian,
\begin{multline}
H_{\rm M1}=-\sqrt{3}\lambda Q^{(1)}(1)-\epsilon Q_{2,0}^{(0)}(0)
-\bm{h}_{\rm s}\cdot\bm{\sigma}-\bm{h}_{\rm o}\cdot\bm{l},
\end{multline}
where the first two terms represent the spin-orbit coupling $\bm{l}\cdot\bm{\sigma}$ and the locking potential with sufficiently large $\epsilon$ that makes the $z$ axis of $p_{z}$ orbital be the principal axis, namely, the crystalline electric field (CEF) potential.
The last two terms are the Zeeman couplings with the spin and orbital angular momenta which represent the molecular fields from the surrounding magnetic moments.
We set $\epsilon=1$ and assume the molecular magnetic field in $zx$ plane, \textit{i.e.}, $\bm{h}_{\rm s,o}=h_{\rm s,o}(\sin\theta,0,\cos\theta)$ in the following discussion.

\begin{figure}[h!]
\centering
\includegraphics[width=7cm]{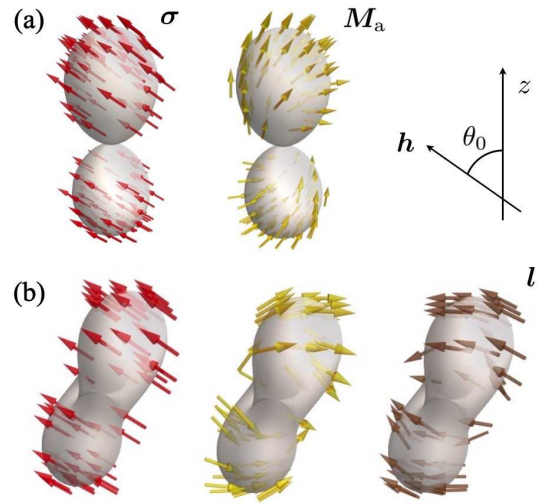}
\caption{Angular distributions of the spin $\bm{\sigma}$, anisotropic M dipole $\bm{M}_{\rm a}$, and orbital angular momentum $\bm{l}$ in the absence of the spin-orbit coupling.
The magnetic field $\bm{h}$ is applied in $zx$ plane, and $\theta$ is fixed at the magic angle $\theta_{0}$.
(a) $h_{\rm o}=0$, (b) $h_{\rm s}=h_{\rm o}=1$.
}
\label{anim}
\end{figure}

First, we consider the case without the spin-orbit coupling, $\lambda=0$, and dominant spin ordering, $h_{\rm s}>0$ and $h_{\rm o}=0$.
The ground state energy is given by $E_{\rm gs}=-2/5-h$, and its eigenstate is
\begin{multline}
\ket{0}=\frac{1}{\sqrt{3}}\left[
\cos\frac{\theta}{2}\Ket{\frac{1}{2},+\frac{1}{2};p}
-\sin\frac{\theta}{2}\Ket{\frac{1}{2},-\frac{1}{2};p}
\right.\\\left.
-\sqrt{2}\cos\frac{\theta}{2}\Ket{\frac{3}{2},+\frac{1}{2};p}
-\sqrt{2}\sin\frac{\theta}{2}\Ket{\frac{3}{2},-\frac{1}{2};p}
\right].
\end{multline}
Note that the wavefunction is independent of the magnitude $h_{\rm s}$.
In this ground state, the charge and three M-dipole distributions are given by
\begin{align}
&
\overline{\rho}(\hat{\bm{r}})=3\hat{z}^{2},
\cr&
\overline{\bm{\sigma}}(\hat{\bm{r}})=3\hat{z}^{2}(\sin\theta, 0, \cos\theta),
\quad
\overline{\bm{l}}(\hat{\bm{r}})=0,
\cr&
\overline{\bm{M}}_{\rm a}(\hat{\bm{r}})=\frac{3\hat{z}\cos\theta}{5\sqrt{10}}(
3\hat{x}-2\hat{z}\tan\theta,
3\hat{y},
4\hat{z}+3\hat{x}\tan\theta
).
\end{align}
The expectation values are obtained by the angular average as
\begin{align}
&
\overline{\rho}=1,
\cr&
\overline{\bm{\sigma}}=(\sin\theta,0,\cos\theta),
\cr&
\overline{\bm{M}}_{\rm a}=-\frac{2}{5\sqrt{10}}(\sin\theta,0,-2\cos\theta).
\end{align}
Note that $\overline{\bm{\sigma}}$ and $\overline{\bm{M}}_{\rm a}$ become perpendicular with each other at the so-called magic angle $\theta_{0}=\cos^{-1}(1/\sqrt{3})\simeq 54.7356^{\circ}$~\cite{Stohr_PhysRevLett.75.3748}.
These angular distributions at $\theta=\theta_{0}$ are shown in Fig.~\ref{anim}(a).

When we switch on the molecular field to $\bm{l}$, the relative directions of three M dipoles are altered.
Figure~\ref{anim}(b) shows three M-dipole angular distributions for $h_{\rm s}=h_{\rm o}=1$ at the magic angle $\theta=\theta_{0}$.
$\bm{l}$ and $\bm{\sigma}$ tend to align in the same direction of the magnetic field because of the Zeeman couplings, while $\bm{M}_{\rm a}$ tends to direct the opposite direction to $\bm{l}$ and $\bm{\sigma}$.
Note that the angular average of the M dipole distributions lies in $zx$ plane.
The shape of the wavefunction is also deformed.
As shown in this example, a careful consideration of the contributions from three distinct M dipoles is necessary in analyzing XMCD spectra.

\begin{figure}[h!]
\centering
\includegraphics[width=7cm]{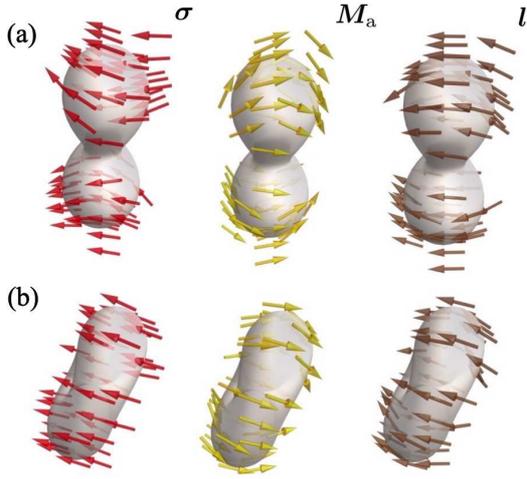}
\caption{Angular distribution of the spin $\bm{\sigma}$, anisotropic M dipole $\bm{M}_{\rm a}$, and orbital angular momentum $\bm{l}$ in the strong spin-orbit coupling limit.
The magnetic field $\bm{h}_{\rm s}$ is applied in $zx$ plane and $\bm{h}_{\rm o}=\bm{0}$.
$\theta$ is fixed at the magic angle $\theta_{0}$.
(a) Under weak magnetic field $h_{\rm s}=0.1$, and (b) under strong magnetic field $h_{\rm s}=1$.
}
\label{anim2}
\end{figure}

Moreover, we discuss the influence of the spin-orbit coupling.
To this end, let us consider the limit of strong spin-orbit coupling, $\lambda=\infty$.
For simplicity, we put $h_{\rm o}=0$.
In this case, we obtain the explicit analytical solution only for $\theta=\theta_{0}$, and the ground state energy is given by
\begin{align}
E_{\rm gs}=-\frac{1}{15}\sqrt{9 + 125 h_{\rm s}^2 + 10h_{\rm s} \sqrt{27 + 100 h_{\rm s}^2}}.
\end{align}
The corresponding eigenstate is also obtained analytically, however we omit it as it is rather lengthy.
The angular distributions of three M dipoles for weak magnetic field $h_{\rm s}=0.1$ and strong one $h_{\rm s}=1$ are shown in Fig.~\ref{anim2}.
The angular average of the M dipole distributions also lies in $zx$ plane.
Interestingly, the effect of the spin-orbit coupling on the mutual interplay among three M dipoles is similar to that shown in Fig.~\ref{anim}(b) in which the molecular fields are applied both on $\bm{l}$ and $\bm{\sigma}$ in the absence of $\lambda$.
This is because the effective coupling among three M dipoles arises through the E quadrupole $Q_{2,0}^{(0)}(0)$.
Namely, there are the 3rd order couplings in the free energy,
\begin{align}
&
(M_{\rm a}^{z})^{2}\left[Q^{(1)}(1)+Q_{2,0}^{(0)}(0)\right],
\quad
l_{z}M_{\rm a}^{z}Q^{(1)}(1),
\quad
\sigma_{z}M_{\rm a}^{z}Q_{2,0}^{(0)}(0).
\end{align}
The combination of these couplings gives rise to the effective coupling among three M dipoles, the spin-orbit coupling, $\bm{l}\cdot\bm{\sigma}$, and the CEF potential, $Q_{2,0}^{(0)}(0)$.
The general treatment of the algebra of multipoles and the Landau expansion in terms of them are discussed in Ref.~\citen{comment_landau_exp} in detail.

\section{Summary}

In this paper, we have derived a whole series of expressions for four species of multipoles in single-centered spinful systems in Eq.~(\ref{sfmpop}).
The matrix elements of these operators in total angular momentum basis are given by Eqs.~(\ref{sfwe}), (\ref{rel1}), and (\ref{rel2}) with use of Eqs.~(\ref{rq}), (\ref{rm}), (\ref{rl12}), and (\ref{p12}).
The multipoles in the expansions of electromagnetic potentials are related to those in the complete set as in Eqs.~(\ref{mpm})--(\ref{mpg}).
We have proposed two complementary methods of visualizations of electronic states, and demonstrated by considering the monopole and toroidal dipole orderings as typical examples.
Moreover, we have discussed the mutual interplay among three distinct magnetic dipoles which are usually active in non $s$-orbital spinful systems.
The obtained complete multipole basis set is useful to describe arbitrary single-centered electron systems including cluster, bond, and momentum-space extensions.

%%% Acknowledge %%%
\begin{acknowledgments}
We would like to thank Yuichi Yamasaki, Hironori Nakao, and Hiroshi Amitsuka for fruitful discussions on XMCD spectra.
This work was supported by JSPS KAKENHI Grants Numbers JP15H05885, JP18H04296 (J-Physics), JP18K13488, JP19K03752, JP19H01834, and JP20J21838.
\end{acknowledgments}

\appendix

\section{Derivation of Reduced Matrix Elements of $X_{l,m}^{\rm (orb)}$}\label{app1}

Here, the derivation of the reduced matrix elements, Eqs.~(\ref{rq}), (\ref{rm}), and (\ref{relmt}) is given.

\subsection{For electric and magnetic multipoles}

Let us begin with the derivation of the reduced matrix element of $Q_{l,m}^{\rm (orb)}$.
We omit the quantum numbers $n_{1}$ and $n_{2}$ for notational simplicity.
It is straightforward to evaluate the matrix element,
\begin{align}
&
\braket{L_{1}M_{1}|Q_{l,m}^{\rm (orb)}|L_{2}M_{2}}
=\braket{r^{l}}_{12}
\sqrt{\frac{4\pi}{2l+1}}
\int d\hat{\bm{r}}\,Y_{L_{1},M_{1}}^{*}Y_{l,m}Y_{L_{2},M_{2}}
\cr&\quad
=(-1)^{M_{1}}\braket{r^{l}}_{12}\sqrt{(2L_{1}+1)(2L_{2}+1)}
\cr&\quad\quad\times
\begin{pmatrix} L_{1} & L_{2} & l \\ 0 & 0 & 0 \end{pmatrix}
\begin{pmatrix} L_{1} & L_{2} & l \\ -M_{1} & M_{2} & m \end{pmatrix},
\label{qmat}
\end{align}
where we have used the integration formula for the product of three spherical harmonics.
By comparing Eqs.~(\ref{qmat}) with Eq.~(\ref{wedecomp}), we obtain Eq.~(\ref{rq}).

Next, we evaluate the matrix element of $M_{l,m}^{\rm (orb)}$.
By using the relations derived from the definition of the vector spherical harmonics, $\bm{\nabla}O_{l,m}=r^{l-1}\sqrt{4\pi l}\bm{Y}_{l,m}^{l-1}$ and $\bm{l}Y_{L_{2},M_{2}}=\sqrt{L_{2}(L_{2}+1)}\bm{Y}_{L_{2},M_{2}}^{L_{2}}$, we obtain
\begin{align}
&
\braket{L_{1}M_{1}|M_{l,m}^{\rm (orb)}|L_{2}M_{2}}
=(-1)^{m}\frac{2\sqrt{4\pi l}\sqrt{L_{2}(L_{2}+1)}\braket{r^{l-1}}_{12}}{l+1}
\cr&\quad\quad\times
\int d\hat{\bm{r}}\,Y_{L_{1},M_{1}}^{*}\bm{Y}_{l,-m}^{l-1*}\cdot\bm{Y}_{L_{2},M_{2}}^{L_{2}}
\cr&\quad
=(-1)^{M_{1}}\sqrt{l(2l+1)(2l-1)(2L_{1}+1)L_{2}(L_{2}+1)}
\cr&\quad\times
\frac{2(2L_{2}+1)\braket{r^{l-1}}_{12}}{l+1}
\begin{pmatrix} L_{1} & L_{2} & l-1 \\ 0 & 0 & 0 \end{pmatrix}
\begin{Bmatrix} l-1 & l & 1 \\ L_{2} & L_{2} & L_{1} \end{Bmatrix}
\cr&\quad\times
\begin{pmatrix} L_{1} & L_{2} & l \\ -M_{1} & M_{2} & m \end{pmatrix},
\end{align}
where we have used the integration formula for the product of vector spherical harmonics.
Then, we obtain Eq.~(\ref{rm}) by the comparison with Eq.~(\ref{wedecomp}).

\subsection{Relation between E, M and MT, ET multipoles}

Moreover, we discuss the relation between the multipoles and toroidal multipoles.
First, let us consider $T_{l,m}^{\rm (orb)}$.
By the relation, $\bm{r}\times(\bm{\nabla}O_{l,m})=i(\bm{l}O_{l,m})$, the following expression appearing in the definition of $T_{l,m}^{\rm (orb)}$ is reexpressed as
\begin{align*}
&
\frac{1}{2}[(\bm{\nabla}O_{l,m})\cdot(\bm{r}\times\bm{l})-(\bm{l}\times\bm{r})\cdot(\bm{\nabla}O_{l,m})]
\cr&\quad\quad
=\frac{1}{2}[-(\bm{r}\times\bm{\nabla}O_{l,m})\cdot\bm{l}-\bm{l}\cdot(\bm{r}\times\bm{\nabla}O_{l,m})]
\cr&\quad\quad
=-\frac{i}{2}\biggl[(\bm{l}O_{l,m})\cdot\bm{l}+\bm{l}\cdot(\bm{l}O_{l,m})\biggr]
=-\frac{i}{2}\left[\bm{l}^{2}O_{l,m}-O_{l,m}\bm{l}^{2}\right],
\end{align*}
where we have used the identities,
\begin{align*}
(\bm{l}O_{l,m})=\bm{l}O_{l,m}-O_{l,m}\bm{l},
\quad\quad
\bm{l}\cdot(\bm{l}O_{l,m})=\bm{l}^{2}O_{l,m}-\bm{l}O_{l,m}\cdot\bm{l}.
\end{align*}
Note that the parenthesis specifies on what range the operators $\bm{l}$ and $\bm{\nabla}$ act.
Acting on the bra (ket) state on the first (second) term in the above expression, we obtain Eq.~(\ref{relmt}) for $T_{l,m}^{\rm (orb)}$ and $Q_{l,m}^{\rm (orb)}$.

Similarly, by comparing both definitions of $G_{l,m}^{\rm (orb)}$ and $M_{l,m}^{\rm (orb)}$, we obtain
\begin{align}
&
G_{l,m}^{\rm (orb)}=\frac{4i}{2(l+1)^{2}(l+2)}[(\bm{\nabla}O_{l,m})\cdot\bm{l}\,\bm{l}^{2}-\bm{l}^{2}\,\bm{l}\cdot(\bm{\nabla}O_{l,m})]
\cr&\quad\quad
=-i\frac{1}{(l+1)(l+2)}\biggl[
\bm{l}^{2}M_{l,m}^{\rm (orb)}-M_{l,m}^{\rm (orb)}\bm{l}^{2}
\biggr].
\end{align}
Thus, the same relation, Eq.~(\ref{relmt}), holds for $G_{l,m}^{\rm (orb)}$ and $M_{l,m}^{\rm (orb)}$.

\section{Derivation of Reduced Matrix Elements of $X_{l,m}^{(s)}(k)$}\label{app2}

In this appendix, the derivation of the reduced matrix elements of the spinful multipole operator $X_{l,m}^{(s)}(k)$ is given.

Let us begin with the Wigner-Eckart theorem, Eq.~(\ref{sfwe}).
By using the completeness relation for $3j$ symbol, we can revert Eq.~(\ref{sfwe}) as
\begin{multline}
\braket{J_{1}||X_{l}^{(s)}(k)||J_{2}}
=\sum_{mM_{1}M_{2}}(-1)^{J_{1}+M_{1}}\braket{J_{1}M_{1}|X_{l,m}^{(s)}(k)|J_{2}M_{2}}
\\\times
\begin{pmatrix} J_{1} & J_{2} & l \\ -M_{1} & M_{2} & m \end{pmatrix}.
\label{revwe}
\end{multline}
From the definition of $X_{l,m}^{(s)}(k)$, Eq.~(\ref{sfmpop}) and by applying the Wigner-Eckart theorem separately  for the orbital and spin parts, we obtain the matrix element in the right-hand side of Eq.~(\ref{revwe}) as
\begin{align*}
&
\braket{J_{1}M_{1}|X_{l,m}^{(s)}(k)|J_{2}M_{2}}
=i^{s+k}\sum_{M_{1}'M_{2}'m'\sigma_{1}\sigma_{2}n}
(-1)^{l+m+J_{1}+M_{1}+J_{2}+M_{2}}
\cr&\hspace{0.5cm}\times
\sqrt{(2l+1)(2J_{1}+1)(2J_{2}+1)}
\braket{L_{1}M_{1}'|X_{l+k,m'}^{({\rm orb})}|L_{2}M_{2}'}
\cr&\hspace{1cm}\times
\braket{\sigma_{1}|\sigma_{s,n}|\sigma_{2}}
\begin{pmatrix} l+k & l & s \\ m' & -m & n \end{pmatrix}
\begin{pmatrix} L_{1} & J_{1} & 1/2 \\ M_{1}' & -M_{1} & \sigma_{1} \end{pmatrix}
\cr&\hspace{1cm}\times
\begin{pmatrix} L_{2} & J_{2} & 1/2 \\ M_{2}' & -M_{2} & \sigma_{2} \end{pmatrix}
\cr&
=
i^{s+k}\sum_{M_{1}'M_{2}'m'\sigma_{1}\sigma_{2}n}
(-1)^{l+m+J_{1}+M_{1}+J_{2}+M_{2}}(-1)^{L_{1}+M_{1}'}(-1)^{s+\sigma_{1}+1/2}
\cr&\hspace{1cm}\times
\sqrt{(2l+1)(2J_{1}+1)(2J_{2}+1)}\sqrt{(1-s)!(2+s)!}
\cr&\hspace{2cm}\times
\braket{L_{1}||X_{l+k}^{({\rm orb})}||L_{2}}
\begin{pmatrix} L_{1} & J_{1} & 1/2 \\ M_{1}' & -M_{1} & \sigma_{1} \end{pmatrix}
\cr&\hspace{2cm}\times
\begin{pmatrix} L_{2} & J_{2} & 1/2 \\ M_{2}' & -M_{2} & \sigma_{2} \end{pmatrix}
\begin{pmatrix} l+k & l & s \\ m' & -m & n \end{pmatrix}
\cr&\hspace{2cm}\times
\begin{pmatrix} L_{1} & L_{2} & l+k \\ -M_{1}' & M_{2}' & m' \end{pmatrix}
\begin{pmatrix} 1/2 & 1/2 & s \\ -\sigma_{1} & \sigma_{2} & n \end{pmatrix}.
\end{align*}
Then, we substitute this expression into Eq.~(\ref{revwe}), and using the definition of $9j$ symbol in terms of the summation of product of six $3j$ symbols, we obtain
\begin{align}
&
\braket{J_{1}||X_{l}^{(s)}(k)||J_{2}}
=
i^{s+k}(-1)^{s}
\braket{L_{1}||X_{l+k}^{({\rm orb})}||L_{2}}
\cr&\quad\times
\sqrt{(2l+1)(2J_{1}+1)(2J_{2}+1)(1-s)!(2+s)!}
\cr&\hspace{1cm}\times
\begin{Bmatrix}
L_{1} & J_{1} & 1/2 \\
L_{2} & J_{2} & 1/2 \\
l+k & l & s
\end{Bmatrix}.
\label{eq1}
\end{align}
This expression is Eq.~(\ref{ssex}) for spin sector ($s=1$).
For charge sector, by putting $s=k=0$ and using the identity,
\begin{align*}
\begin{Bmatrix}
L_{1} & J_{1} & 1/2 \\
L_{2} & J_{2} & 1/2 \\
l & l & 0
\end{Bmatrix}
=\frac{(-1)^{J_{1}+1/2+L_{2}+l}}{\sqrt{2(2l+1)}}
\begin{Bmatrix}
L_{1} & L_{2} & l \\
J_{2} & J_{1} & 1/2
\end{Bmatrix},
\end{align*}
we obtain Eq.~(\ref{csex}).

\section{Relation between Multipoles through Potential Expansion and Spinful Multipoles}\label{app3}

In this appendix, we express the multipoles $X_{l,m}$ defined through the electromagnetic potentials in Eqs.~(\ref{pphi}) and (\ref{pa}) in terms of spinful multipole basis set, Eq.~(\ref{sfmpop}).
Since $X_{l,m}$ always contains the orbital part, we discuss the remaining part $\delta X_{l,m}=X_{l,m}-X_{l,m}^{(0)}(0)$ containing the spin operator $\bm{\sigma}$.

Let us begin with $\delta M_{l,m}=\bm{\sigma}\cdot(\bm{\nabla}O_{l,m})$.
Since $\bm{\nabla}O_{l,m}=r^{l-1}\sqrt{4\pi l}\bm{Y}_{l,m}^{l-1}$ and by comparing the definitions of $M_{l,m}^{(1)}(-1)$ with that of the vector spherical harmonics,
\begin{align}
\bm{Y}_{l,m}^{l+k}(\hat{\bm{r}})=(-1)^{l+m}\sqrt{2l+1}\sum_{n}
\begin{pmatrix} l+k & l & 1 \\ m-n & -m & n \end{pmatrix}Y_{l',m-n}(\hat{\bm{r}})\bm{e}_{1,n},
\label{vsh}
\end{align}
we obtain Eq.~(\ref{mpm}) as
\begin{align}
\delta M_{l,m}=\bm{\sigma}\cdot(\bm{\nabla}O_{l,m})=\sqrt{l(2l-1)}M_{l,m}^{(1)}(-1).
\end{align}

For the MT multipole, $\delta T_{l,m}$ is proportional to $(\bm{r}\times\bm{\sigma})\cdot(\bm{\nabla}O_{l,m})$, which can be deformed as $(\bm{r}\times\bm{\sigma})\cdot(\bm{\nabla}O_{l,m})=-(\bm{r}\times\bm{\nabla}O_{l,m})\cdot\bm{\sigma}=-i(\bm{l}O_{l,m})\cdot\bm{\sigma}$.
With $\bm{l}Y_{l,m}=\sqrt{l(l+1)}\bm{Y}_{l,m}^{l}$ and the comparison of the definitions between $\bm{Y}_{l,m}^{l}$ and $T_{l,m}^{(1)}(0)$, we have the relation, $\bm{\sigma}\cdot(\bm{l}O_{l,m})=-i\sqrt{l(l+1)}T_{l,m}^{(1)}(0)$.
Thus, we obtain Eq.~(\ref{mpt}) as
\begin{align}
\delta T_{l,m}=-\sqrt{\frac{l}{l+1}}T_{l,m}^{(1)}(0).
\end{align}

As for the ET multipole, first we consider the first part of Eq.~(\ref{gfac}).
This part can be deformed as
\begin{align*}
&
\frac{1}{2}\sum_{ij}[(\nabla_{i}\nabla_{j}O_{l,m})(\bm{r}\times\bm{l})_{i}-(\bm{l}\times\bm{r})_{i}(\nabla_{i}\nabla_{j}O_{l,m})]\sigma_{j}
\cr&\quad
=-\frac{1}{2}\sum_{ij}\biggl[
[(\bm{r}\times\bm{\nabla})_{i}(\nabla_{j}O_{l,m})]l_{i}+l_{i}[(\bm{r}\times\bm{\nabla})_{i}\nabla_{j}O_{l,m}]
\biggr]\sigma_{j}
\cr&\quad
=-\frac{i}{2}\sum_{ij}\biggl[
(l_{i}\nabla_{j}O_{l,m})l_{i}+l_{i}(l_{i}\nabla_{j}O_{l,m})
\biggr]\sigma_{j}.
\end{align*}
By using $(l_{i}\nabla_{j}O_{l,m})=l_{i}(\nabla_{j}O_{l,m})-(\nabla_{j}O_{l,m})l_{i}$, we have
\begin{align*}
=-\frac{i}{2}\biggl[
\bm{l}^{2}(\bm{\sigma}\cdot\bm{\nabla}O_{l,m})-(\bm{\sigma}\cdot\bm{\nabla}O_{l,m})\bm{l}^{2}
\biggr].
\end{align*}
Then, the first term becomes
\begin{align*}
\delta G_{l,m}^{\rm (1st)}=
-i\frac{\sqrt{l(2l-1)}}{(l+1)(l+2)}
\biggl[
\bm{l}^{2}M_{l,m}^{(1)}(-1)-M_{l,m}^{(1)}(-1)\bm{l}^{2}
\biggr].
\end{align*}
By noticing the identity derived from Eqs.~(\ref{rel1}) and (\ref{rel2}),
\begin{multline*}
i\,R_{l-1}(L_{1},L_{2})\braket{J_{1}M_{1}|M_{l,m}^{(1)}(-1)|J_{2}M_{2}}
\\
=\braket{J_{1}M_{1}|G_{l,m}^{(1)}(-1)|J_{2}M_{2}},
\end{multline*}
we obitain
\begin{align}
\delta G_{l,m}^{\rm (1st)}=
\frac{l\sqrt{l(2l-1)}}{l+2}G_{l,m}^{(1)}(-1).
\end{align}

Next, we consider the second part of Eq.~(\ref{gfac}), which is deformed as
\begin{align*}
&
\frac{1}{2}\sum_{ij}\biggl[(\bm{r}\times\bm{\sigma})_{i}(\nabla_{i}\nabla_{j}O_{l,m})l_{j}
+l_{j}(\bm{r}\times\bm{\sigma})_{i}
(\nabla_{i}\nabla_{j}O_{l,m})\biggr]
\cr&\quad
=
-\frac{i}{2}\sum_{ij}\biggl[(l_{i}\nabla_{j}O_{l,m})l_{j}
+l_{j}(l_{i}\nabla_{j}O_{l,m})\biggr]\sigma_{i}.
\end{align*}
Meanwhile, the definition of $G_{l,m}^{(1)}(0)$ can be deformed as
\begin{align*}
&
\frac{l+1}{2}G_{l,m}^{(1)}(0)
=i\sum_{m'}\sum_{n}(-1)^{l+m}\sqrt{2l+1}
\begin{pmatrix} l & l & 1 \\ m' & -m & n \end{pmatrix}
\cr&\quad\quad\quad\times
\frac{\bm{l}\cdot(\bm{\nabla}O_{l,m'})+(\bm{\nabla}O_{l,m'})\cdot\bm{l}}{2}\sigma_{1,n}
\cr&
=i\sum_{i}\sum_{m'}\sum_{n}(-1)^{l+m}\sqrt{2l+1}
\begin{pmatrix} l & l & 1 \\ m' & -m & n \end{pmatrix}
\cr&\quad\quad\quad\times
\frac{l_{i}(\nabla_{i}O_{l,m'})+(\nabla_{i}O_{l,m'})l_{i}}{2}\bm{e}_{1,n}\cdot\bm{\sigma}
\cr&
=\sqrt{\frac{4\pi}{2l+1}}i\sum_{i}\sum_{m'}\sum_{n}(-1)^{l+m}\sqrt{2l+1}
\begin{pmatrix} l & l & 1 \\ m' & -m & n \end{pmatrix}
\cr&\quad\quad\quad\times
\frac{l_{i}(\nabla_{i}r^{l}Y_{l,m'})+(\nabla_{i}r^{l}Y_{l,m'})l_{i}}{2}\bm{e}_{1,n}\cdot\bm{\sigma}
\cr&
=\sqrt{\frac{4\pi}{2l+1}}i
\sum_{ij}\frac{l_{i}\sigma_{j}(\nabla_{i}r^{l}Y_{l,m\,j}^{l})+(\nabla_{j}r^{l}Y_{l,m\,j}^{l})\sigma_{i}l_{j}}{2}
\cr&
=\frac{i}{2\sqrt{l(l+1)}}
\sum_{ij}\biggl[l_{j}(l_{i}\nabla_{j}O_{l,m})+(l_{i}\nabla_{j}O_{l,m})l_{j}\biggr]\sigma_{i}.
\end{align*}
Therefore, we have
\begin{multline*}
-\frac{i}{2}\sum_{ij}\biggl[
l_{j}(l_{i}\nabla_{j}O_{l,m})
+(l_{i}\nabla_{j}O_{l,m})l_{j}
\biggr]\sigma_{i}
\\
=
-\frac{l+1}{2}\sqrt{l(l+1)}G_{l,m}^{(1)}(0),
\quad\quad
\end{multline*}
and the second term becomes
\begin{align}
\delta G_{l,m}^{\rm (2nd)}=-\sqrt{\frac{l}{l+1}}G_{l,m}^{(1)}(0).
\end{align}
Finally, we obtain the relation as in Eq.~(\ref{mpg}).

%\bibliographystyle{jpsj}
%\bibliography{ref}

\newpage

\onecolumn
\setcounter{section}{23}

\begin{center}\bf\large
Supplemental Materials
\end{center}
\begin{quote}
We here provide the algebra of multipoles, and the Landau expansion are derived in order to discuss mutual coupling among multipoles.
Then, the matrix elements of all active multipoles in $J=1/2$ and $J=3/2$ with $L=0$, $1$ used in the main text.
Moreover, the irreducible basis functions of cubic $O_{h}$ and hexagonal $D_{6h}$ groups are also given.
The expressions of the multipole operators and their matrix elements in 32 crystallographic point groups are obtained by using the coefficient of linear combination of the tesseral basis functions.
\end{quote}

\section{Algebra and Coupling of Multipoles}

\subsection{The property of multipole operators}

Let us consider a complete multipole basis set, $\set{X_{i}}$.
The multipole operator $X_{i}$ is $d\times d$ matrix and hermite, $X_{i}=X_{i}^{\dagger}$.
There are $d^{2}$ mutually independent matrices, and they satisfy
\begin{align}
{\rm Tr}(X_{i}X_{j})=d\,\delta_{ij}
\quad
(i,j=0,1,2,\cdots d^{2}-1).
\label{eq:orth}
\end{align}
The $0$-th component ($i=0$) represents the unit matrix, $X_{0}\equiv I$, and the other matrices are traceless ${\rm Tr}(X_{i})=0$ ($i\ne0$).

\subsection{The matrix product}

Since $\set{X_{i}}$ constitute a complete set, we expand a product of multipoles as
\begin{align}
X_{i}X_{j}=\sum_{k}h_{ijk}X_{k},
\label{eq:exp}
\end{align}
where the coefficient $h_{ijk}$ is complex in general.
By this relation and eq.~(\ref{eq:orth}), we obtain
\begin{align}
h_{ijk}=\frac{1}{d}{\rm Tr}(X_{i}X_{j}X_{k}).
\end{align}
Thus, the cyclic permutation of $h_{ijk}$ is identical, \textit{i.e.}, $h_{ijk}=h_{jki}=h_{kij}$.
Moreover, taking hermite conjugate of eq.~(\ref{eq:exp}) gives
\begin{align}
X_{j}X_{i}=\sum_{k}h_{ijk}^{*}X_{k}
\quad\Rightarrow\quad
h_{ijk}=h_{jik}^{*}.
\end{align}
When we decompose $h_{ijk}$ into real and imaginary parts as $h_{ijk}=g_{ijk}+if_{ijk}$, each part satisfies
\begin{align*}
g_{ijk}=g_{jik},
\quad
f_{ijk}=-f_{jik}.
\end{align*}
Together with the property of $h_{ijk}$ w.r.t. the cyclic permutation, we show that $g_{ijk}$ is completely symmetric, and $f_{ijk}$ is completely antisymmetric w.r.t. any permutations of a pair of indices,
\begin{align}
&
g_{ijk}=g_{jki}=g_{kij}=g_{jik}=g_{ikj}=g_{kji},
\\&
f_{ijk}=f_{jki}=f_{kij}=-f_{jik}=-f_{ikj}=-f_{kji}.
\end{align}
Therefore,
\begin{align}
X_{i}X_{j}=\sum_{k=0}^{d^{2}-1}(g_{ijk}+if_{ijk})X_{k}.
\end{align}
By tracing the above relation, we obtain
\begin{align}
d\,\delta_{ij}=(g_{ij0}+if_{ij0})d.
\end{align}
Namely, we have
\begin{align}
g_{ij0}=\delta_{ij},
\quad
f_{ij0}=0.
\end{align}

By using eq.~(\ref{eq:exp}) and its hermite conjugate, we obtain
\begin{align}
g_{ijk}=\frac{1}{2d}{\rm Tr}\biggl([X_{i},X_{j}]_{+}X_{k}\biggr),
\quad
f_{ijk}=\frac{1}{2di}{\rm Tr}\biggl([X_{i},X_{j}]_{-}X_{k}\biggr),
\label{stfac}
\end{align}
where $[A,B]_{\pm}=AB\pm BA$.

\subsection{Free energy and Landau expansion}

\subsubsection{The free energy in the mean-field approximation}

Let us consider a generalized exchange Hamiltonian,
\begin{align}
H=-\frac{1}{2}\sum_{ss'}\sum_{ij}'D_{ij}^{ss'}X^{s}_{i}X^{s'}_{j},
\end{align}
where $s$ represents a site of the operators.
We assume that $D_{ij}^{ss'}=D_{ji}^{s's}$, $D_{ii}^{ss'}=D_{ij}^{ss}=0$.
$\displaystyle\sum_{i}'=\sum_{i}^{\ne0}$ indicates that the summation excludes $i=0$ component.

We consider the one-body trial Hamiltonian, $\displaystyle H_{0}(\phi)=-\sum_{s}\sum_{i}'\phi_{i}^{s}X_{i}^{s}\equiv\sum_{s}H_{0s}(\phi^{s})$, and adopt the Feynman's variational principle, we obtain the trial free energy as
\begin{align}
F(\phi)=\sum_{s}F_{0s}(\phi^{s})-\frac{1}{2}\sum_{ss'}\sum_{ij}'D_{ij}^{ss'}\braket{X_{i}^{s}}\braket{X_{j}^{s'}}+\sum_{s}\sum_{i}'\phi_{i}^{s}\braket{X_{i}^{s}},
\quad
F_{0s}(\phi^{s})=-\frac{1}{\beta}\ln{\rm Tr}\left(e^{-\beta H_{0s}}\right).
\label{eq:fe}
\end{align}
Here, the expectation value w.r.t. the trial Hamiltonian is given by
\begin{align}
\braket{X_{i}^{s}}=\frac{{\rm Tr}(e^{-\beta H_{0s}}X_{i}^{s})}{{\rm Tr}(e^{-\beta H_{0s}})}
=-\frac{\partial F_{0s}}{\partial\phi_{i}^{s}}
\quad
(i\ne0).
\end{align}
The variational condition of $F(\phi)$ gives
\begin{align}
\phi_{i}^{s}=\sum_{s'}\sum_{j}'D_{ij}^{ss'}\braket{X_{j}^{s'}}
\quad
(i\ne 0).
\end{align}
This is nothing but the self-consistent equation in the mean-field approximation.

\subsubsection{Landau expansion}

Let us expand the free energy, $F_{0s}(\phi^{s})$, with respect to $\phi_{i}^{s}$.
Using the relations,
\begin{align*}
&
\frac{1}{d}{\rm Tr}(X_{i}^{s}X_{j}^{s}X_{k}^{s})=h_{ijk},
\quad
\frac{1}{d}{\rm Tr}(X_{i}^{s}X_{j}^{s}X_{k}^{s}X_{l}^{s})=\sum_{m}h_{ijm}h_{mkl},
\end{align*}
we obtain
\begin{align*}
\frac{1}{d}{\rm Tr}\left(e^{-\beta H_{0s}}\right)
=1+\frac{\beta^{2}}{2}\sum_{i}'\phi_{i}^{s}\phi_{i}^{s}
+\frac{\beta^{3}}{3!}\sum_{ijk}'h_{ijk}\phi_{i}^{s}\phi_{j}^{s}\phi_{k}^{s}+\frac{\beta^{4}}{4!}\sum_{ijkl}'\sum_{m}h_{ijm}h_{mkl}\phi_{i}^{s}\phi_{j}^{s}\phi_{k}^{s}\phi_{l}^{s}+\cdots.
\end{align*}
Therefore, the free energy is expanded as
\begin{align*}
F_{0s}(\phi^{s})=-\frac{1}{\beta}\ln d-\frac{\beta}{2}\sum_{i}'\phi_{i}^{s}\phi_{i}^{s}-\frac{\beta^{2}}{3!}\sum_{ijk}'h_{ijk}\phi_{i}^{s}\phi_{j}^{s}\phi_{k}^{s}-\frac{\beta^{3}}{4!}\sum_{ijkl}'\left[\sum_{m}h_{ijm}h_{mkl}-3\delta_{ij}\delta_{kl}\right]\phi_{i}^{s}\phi_{j}^{s}\phi_{k}^{s}\phi_{l}^{s}+\cdots.
\end{align*}
In this expression, as $\phi_{i}^{s}\phi_{j}^{s}\phi_{k}^{s}$ is completely symmetric, the contributions of $f_{ijk}$ in 3rd order vanish, and only the contributions from $g_{ijk}$ remain.
Similarly, since $\phi_{i}^{s}\phi_{j}^{s}$ or $\phi_{k}^{s}\phi_{l}^{s}$ is symmetric with the permutation of $(i,j)$ or $(k,l)$, there remains $\sum_{m}g_{ijm}g_{mkl}$ in $\sum_{m}h_{ijm}h_{mkl}$ in 4th order.
Therefore,
\begin{align}
F_{0s}(\phi^{s})=-\frac{1}{\beta}\ln d-\frac{\beta}{2}\sum_{i}'\phi_{i}^{s}\phi_{i}^{s}-\frac{\beta^{2}}{3!}\sum_{ijk}'g_{ijk}\phi_{i}^{s}\phi_{j}^{s}\phi_{k}^{s}-\frac{\beta^{3}}{4!}\sum_{ijkl}'\epsilon_{ij;kl}\phi_{i}^{s}\phi_{j}^{s}\phi_{k}^{s}\phi_{l}^{s}+\cdots.
\label{eq:fe0}
\end{align}
Here, we have introduced
\begin{align}
\epsilon_{ij;kl}=\sum_{m}g_{ijm}g_{mkl}-3\delta_{ij}\delta_{kl}=\sum_{m}'g_{ijm}g_{mkl}-2\delta_{ij}\delta_{kl}.
\end{align}

By the expansion of $F_{0s}(\phi^{s})$ and appropriate replacement of dummy indices, we obtain the relation,
\begin{align}
\braket{X_{i}^{s}}
=\beta \phi_{i}^{s}+\frac{\beta^{2}}{2}\sum_{jk}'g_{ijk}\phi_{j}^{s}\phi_{k}^{s}+\frac{\beta^{3}}{3!}\sum_{jkl}'\epsilon_{ij;kl}\phi_{j}^{s}\phi_{k}^{s}\phi_{l}^{s}+\cdots.
\label{eq:xexp}
\end{align}
In order to invert this relation for $\phi_{i}^{s}$, we assume the relation,
\[
\beta \phi_{i}^{s}
=
\braket{X_{i}^{s}}
-\frac{\beta^{2}}{2}\sum_{jk}'g_{ijk}\phi_{j}^{s}\phi_{k}^{s}-\frac{\beta^{3}}{3!}\sum_{jkl}'\epsilon_{ij;kl}\phi_{j}^{s}\phi_{k}^{s}\phi_{l}^{s}+\cdots,
\]
and insert it iteratively, then we obtain
\begin{align}
\beta \phi_{i}^{s}
&=
\braket{X_{i}^{s}}
-\frac{1}{2}\sum_{jk}'g_{ijk}
\left[\braket{X_{j}^{s}}
-\frac{\beta^{2}}{2}\sum_{j'k'}'g_{jj'k'}\phi_{j'}^{s}\phi_{k'}^{s}\right]
\left[\braket{X_{k}^{s}}
-\frac{\beta^{2}}{2}\sum_{lm}'g_{klm}\phi_{l}^{s}\phi_{m}^{s}\right]
\cr&\quad
-\frac{1}{3!}\sum_{jkl}'\epsilon_{ij;kl}\braket{X_{j}^{s}}\braket{X_{k}^{s}}\braket{X_{l}^{s}}+\cdots
\cr&
=
\braket{X_{i}^{s}}
-\frac{1}{2}\sum_{jk}'g_{ijk}\braket{X_{j}^{s}}\braket{X_{k}^{s}}
+\frac{1}{3}\sum_{jkl}'L_{ij;kl}\braket{X_{j}^{s}}\braket{X_{k}^{s}}\braket{X_{l}^{s}}
+\cdots,
\end{align}
where we have introduced
\[
L_{ij;kl}=\frac{1}{2}\left[3\sum_{m}'g_{ijm}g_{mkl}-\epsilon_{ij;kl}\right]
=\sum_{m}'g_{ijm}g_{mkl}+\delta_{ij}\delta_{kl}=\sum_{m}g_{ijm}g_{mkl}.
\]
Inserting this expression into eq.~(\ref{eq:fe}), we finally obtain the Landau expansion of the free energy
\begin{multline}
F(\braket{X})
=-NT\ln d
+\frac{1}{2}\sum_{ss'}\sum_{ij}'\left[T\delta_{ss'}\delta_{ij}-D_{ij}^{ss'}\right]\braket{X_{i}^{s}}\braket{X_{j}^{s'}}
\\
-\frac{T}{3!}\sum_{s}\sum_{ijk}'g_{ijk}\braket{X_{i}^{s}}\braket{X_{j}^{s}}\braket{X_{k}^{s}}
+\frac{2T}{4!}\sum_{s}\sum_{ijkl}'L_{ij;kl}\braket{X_{i}^{s}}\braket{X_{j}^{s}}\braket{X_{k}^{s}}\braket{X_{l}^{s}}+\cdots,
\end{multline}
where $N$ is the number of the sites.
Note that the lowest order coupling is given by $g_{ijk}$, \textit{i.e.}, eq.~(\ref{stfac}).

The coupling among multipoles is also obtained by the quantity,
\begin{align}
{\rm Tr}(e^{-\beta H_{0s}}X^{s}_{i})=\sum_{k=0}^{\infty}\frac{(-\beta)^{k}}{k!}{\rm Tr}(H_{0s}^{k}X_{i}^{s}),
\end{align}
which becomes finite if the coupling to $X_{i}^{s}$ exists.

\section{$J=1/2$ and $3/2$ with $L=0$ and $1$ System}
\subsection{Basis}
The total angular-momentum basis $\ket{J,M;L}$ used in the main text is given by
\begin{align*}
&
\Ket{\frac{1}{2},+\frac{1}{2};s}=\ket{0,0}\ket{\uparrow},
\quad
\Ket{\frac{1}{2},-\frac{1}{2};s}=\ket{0,0}\ket{\downarrow},
\\&
\Ket{\frac{1}{2},+\frac{1}{2};p}=\sqrt{\frac{2}{3}}\ket{1,+1}\ket{\downarrow}-\sqrt{\frac{1}{3}}\ket{1,0}\ket{\downarrow},
\quad
\Ket{\frac{1}{2},-\frac{1}{2};p}=-\sqrt{\frac{2}{3}}\ket{1,-1}\ket{\uparrow}+\sqrt{\frac{1}{3}}\ket{1,0}\ket{\uparrow},
\\&
\Ket{\frac{3}{2},+\frac{3}{2};p}=\ket{1,+1}\ket{\uparrow},
\quad
\Ket{\frac{3}{2},+\frac{1}{2};p}=\frac{1}{\sqrt{3}}\ket{1,+1}\ket{\downarrow}+\sqrt{\frac{2}{3}}\ket{1,0}\ket{\uparrow},
\cr&
\Ket{\frac{3}{2},-\frac{1}{2};p}=\frac{1}{\sqrt{3}}\ket{1,-1}\ket{\uparrow}+\sqrt{\frac{2}{3}}\ket{1,0}\ket{\downarrow},
\quad
\Ket{\frac{3}{2},-\frac{3}{2};p}=\ket{1,-1}\ket{\downarrow}.
\end{align*}

\subsection{Symmetry of multipoles}
The angle dependence of the multipoles used in the main text is given by
\begin{align*}
&
(x,y,z):\quad x,y,z
\\&
(u,v,yz,zx,xy):\quad \frac{1}{2}(3z^{2}-r^{2}),
\frac{\sqrt{3}}{2}(x^{2}-y^{2}),
\sqrt{3}yz,
\sqrt{3}zx,
\sqrt{3}xy
\\&
(xyz,\alpha x,\alpha y,\alpha z,\beta x,\beta y,\beta z):
\cr&\hspace{2cm}
\sqrt{15}xyz,
\frac{1}{2}x(5x^{2}-3r^{2}),
\frac{1}{2}y(5y^{2}-3r^{2}),
\frac{1}{2}z(5z^{2}-3r^{2})
\cr&\hspace{2cm}
\frac{\sqrt{15}}{2}x(y^{2}-z^{2}),
\frac{\sqrt{15}}{2}y(z^{2}-x^{2}),
\frac{\sqrt{15}}{2}z(x^{2}-y^{2})
\end{align*}

%\newpage
\subsection{Matrix Element}

\small

\subsubsection{Electric multipole (orbital)}

\textbf{rank 0}

$\displaystyle Q{}_{}^{(0)}(0) =
\displaystyle
$

\subsubsection{Electric multipole (spin)}

\textbf{rank 0}

$\displaystyle Q{}_{}^{(1)}(1) =
\displaystyle %
$

\subsection{Magnetic multipole (orbital)}

\textbf{rank 1}

$\displaystyle M{}_{x}^{(0)}(0) =
\displaystyle %
$

\subsection{Magnetic multipole (spin)}

\textbf{rank 0}

$\displaystyle M{}_{}^{(1)}(1) =
\displaystyle %
$

\subsection{Magnetic toroidal multipole (orbital)}

\textbf{rank 1}

$\displaystyle T{}_{x}^{(0)}(0) =
\displaystyle %
$

\subsection{Magnetic toroidal multipole (spin)}

\textbf{rank 1}

$\displaystyle T{}_{x}^{(1)}(0) =
\displaystyle %
$

\subsection{Electric toroidal multipole (spin)}

\textbf{rank 0}

$\displaystyle G{}_{}^{(1)}(1) =
\displaystyle %
$

\normalsize

\section{Cubic and Hexagonal Harmonics}

We introduce the tesseral harmonics as
\begin{align}
&
O_{l,m}=\sqrt{\frac{4\pi}{2l+1}}r^{l}Y_{l,m}=(-1)^{m}O_{l,-m}^{*},
\\&
C_{l,0}=O_{l,0},
\quad
C_{l,m}=\frac{(-1)^{m}}{\sqrt{2}}(O_{l,m}+O_{l,m}^{*}),
\quad
S_{l,m}=\frac{(-1)^{m}}{\sqrt{2}i}(O_{l,m}-O_{l,m}^{*}).
\end{align}
Then, the cubic $O_{h}$ and hexagonal $D_{6h}$ harmonics are expressed by the linear combinations of the tesseral harmonics.
The expressions for other point group are obtained by using the compatibility relation of the irreducible representations.

In the following tables, we abbreviate $C_{lm}\to C[m]$ and $S_{lm}\to S[m]$.
We use the multiplicity label for the basis functions belonging to the same irreducible representation in which $m$ is descending order except for $C[0]$.
This convention sometimes differs from those used in literatures.
Note that the basis functions are orthogonal with each other.

In cubic point group, $(x,y,z)$ components in $T$ and $(u,v)$ components in $E$ are determined so that the matrix elements of $C_{3[111]}$, $C_{2[110]}$, $\sigma_{[110]}$, $C_{4z}$, and $IC_{4z}$ are identical among the same irreducible representaions.

In hexagonal group, we use the principal axes as
\begin{align}
\bm{a}=\bm{e}_{x},
\quad
\bm{b}=-\frac{1}{2}\bm{e}_{x}+\frac{\sqrt{3}}{2}\bm{e}_{y},
\quad
\bm{c}=\bm{e}_{z}.
\end{align}
We take $\bm{e}_{y}$ axis as $C_{2}'$ rotation axis.
$(x,y)$ components in $E$ are determined so that the matrix elements of $C_{3z}$, $C_{2x}$, $C_{2y}$, $\sigma_{x}$, and $\sigma_{y}$ are identical among the same irreducible representations.

%\newpage
\subsection{Cubic multipoles up to rank 11}

\begin{center}
\renewcommand{\arraystretch}{1.1}
\begin{tabular}{ccc}
\textbf{rank 0}\\
\toprule
irrep. & symbol & definition \\
\midrule
$A_{1g}$ &  $Q_{0}^{}$ &     $C[0]$ \\
\bottomrule
\end{tabular}
\hspace{1cm}\vspace{5mm}
\begin{tabular}{ccc}
\textbf{rank 1}\\
\toprule
irrep. & symbol & definition \\
\midrule
$T_{1u}$ &  $Q_{x}^{}$ &     $C[1]$ \\
&  $Q_{y}^{}$ &     $S[1]$ \\
&  $Q_{z}^{}$ &     $C[0]$ \\
\bottomrule
\end{tabular}
\hspace{1cm}\vspace{5mm}
\begin{tabular}{ccc}
\textbf{rank 2}\\
\toprule
irrep. & symbol & definition \\
\midrule
$E_{g}$  &   $Q_{u}^{}$ &     $C[0]$ \\
&   $Q_{v}^{}$ &     $C[2]$ \\ \hline
$T_{2g}$ &  $Q_{yz}^{}$ &     $S[1]$ \\
&  $Q_{zx}^{}$ &     $C[1]$ \\
&  $Q_{xy}^{}$ &     $S[2]$ \\
\bottomrule
\end{tabular}
\hspace{1cm}\vspace{5mm}
\begin{tabular}{ccc}
\textbf{rank 3}\\
\toprule
irrep. & symbol & definition \\
\midrule
$A_{2u}$ & $Q_{3}^{\beta}$ & $S[2]$ \\ \hline
$T_{1u}$ & $Q_{3x}^{\alpha}$ & $- \frac{\sqrt{6} C[1]}{4} + \frac{\sqrt{10} C[3]}{4}$ \\ 
 & $Q_{3y}^{\alpha}$ & $- \frac{\sqrt{6} S[1]}{4} - \frac{\sqrt{10} S[3]}{4}$ \\ 
 & $Q_{3z}^{\alpha}$ & $C[0]$ \\ \hline
$T_{2u}$ & $Q_{3x}^{\beta}$ & $- \frac{\sqrt{10} C[1]}{4} - \frac{\sqrt{6} C[3]}{4}$ \\ 
 & $Q_{3y}^{\beta}$ & $\frac{\sqrt{10} S[1]}{4} - \frac{\sqrt{6} S[3]}{4}$ \\ 
 & $Q_{3z}^{\beta}$ & $C[2]$  \\
\bottomrule
\end{tabular}
\hspace{1cm}\vspace{5mm}
\begin{tabular}{ccc}
\textbf{rank 4}\\
\toprule
irrep. & symbol & definition \\
\midrule
$A_{1g}$ & $Q_{4}^{}$ & $\frac{\sqrt{21} C[0]}{6} + \frac{\sqrt{15} C[4]}{6}$ \\ \hline
$E_{g}$ & $Q_{4u}^{}$ & $\frac{\sqrt{15} C[0]}{6} - \frac{\sqrt{21} C[4]}{6}$ \\ 
 & $Q_{4v}^{}$ & $- C[2]$ \\ \hline
$T_{1g}$ & $Q_{4x}^{\alpha}$ & $- \frac{\sqrt{14} S[1]}{4} - \frac{\sqrt{2} S[3]}{4}$ \\ 
 & $Q_{4y}^{\alpha}$ & $\frac{\sqrt{14} C[1]}{4} - \frac{\sqrt{2} C[3]}{4}$ \\ 
 & $Q_{4z}^{\alpha}$ & $S[4]$ \\ \hline
$T_{2g}$ & $Q_{4x}^{\beta}$ & $- \frac{\sqrt{2} S[1]}{4} + \frac{\sqrt{14} S[3]}{4}$ \\ 
 & $Q_{4y}^{\beta}$ & $- \frac{\sqrt{2} C[1]}{4} - \frac{\sqrt{14} C[3]}{4}$ \\ 
 & $Q_{4z}^{\beta}$ & $S[2]$ \\
\bottomrule
\end{tabular}
\hspace{1cm}\vspace{5mm}
\begin{tabular}{ccc}
\textbf{rank 5}\\
\toprule
irrep. & symbol & definition \\
\midrule
$E_{u}$ & $Q_{5u}^{}$ & $S[4]$ \\ 
 & $Q_{5v}^{}$ & $- S[2]$ \\ \hline
$T_{1u}$ & $Q_{5x}^{\alpha1}$ & $\frac{\sqrt{15} C[1]}{8} - \frac{\sqrt{70} C[3]}{16} + \frac{3 \sqrt{14} C[5]}{16}$ \\ 
 & $Q_{5y}^{\alpha1}$ & $\frac{\sqrt{15} S[1]}{8} + \frac{\sqrt{70} S[3]}{16} + \frac{3 \sqrt{14} S[5]}{16}$ \\ 
 & $Q_{5z}^{\alpha1}$ & $C[0]$ \\ \hline
$T_{1u}$ & $Q_{5x}^{\alpha2}$ & $\frac{\sqrt{21} C[1]}{8} + \frac{9 \sqrt{2} C[3]}{16} + \frac{\sqrt{10} C[5]}{16}$ \\ 
 & $Q_{5y}^{\alpha2}$ & $\frac{\sqrt{21} S[1]}{8} - \frac{9 \sqrt{2} S[3]}{16} + \frac{\sqrt{10} S[5]}{16}$ \\ 
 & $Q_{5z}^{\alpha2}$ & $C[4]$ \\ \hline
$T_{2u}$ & $Q_{5x}^{\beta}$ & $\frac{\sqrt{7} C[1]}{4} - \frac{\sqrt{6} C[3]}{8} - \frac{\sqrt{30} C[5]}{8}$ \\ 
 & $Q_{5y}^{\beta}$ & $- \frac{\sqrt{7} S[1]}{4} - \frac{\sqrt{6} S[3]}{8} + \frac{\sqrt{30} S[5]}{8}$ \\ 
 & $Q_{5z}^{\beta}$ & $C[2]$ \\ 
\bottomrule
\end{tabular}
\hspace{1cm}\vspace{5mm}
\begin{tabular}{ccc}
\textbf{rank 6}\\
\toprule
irrep. & symbol & definition \\
\midrule
$A_{1g}$ & $Q_{6}^{}$ & $\frac{\sqrt{2} C[0]}{4} - \frac{\sqrt{14} C[4]}{4}$ \\ \hline
$A_{2g}$ & $Q_{6}^{\beta}$ & $\frac{\sqrt{11} C[2]}{4} - \frac{\sqrt{5} C[6]}{4}$ \\ \hline
$E_{g}$ & $Q_{6u}^{}$ & $\frac{\sqrt{14} C[0]}{4} + \frac{\sqrt{2} C[4]}{4}$ \\ 
 & $Q_{6v}^{}$ & $\frac{\sqrt{5} C[2]}{4} + \frac{\sqrt{11} C[6]}{4}$ \\ \hline
$T_{1g}$ & $Q_{6x}^{\alpha}$ & $\frac{\sqrt{3} S[1]}{4} - \frac{\sqrt{30} S[3]}{8} - \frac{\sqrt{22} S[5]}{8}$ \\ 
 & $Q_{6y}^{\alpha}$ & $- \frac{\sqrt{3} C[1]}{4} - \frac{\sqrt{30} C[3]}{8} + \frac{\sqrt{22} C[5]}{8}$ \\ 
 & $Q_{6z}^{\alpha}$ & $S[4]$ \\ \hline
$T_{2g}$ & $Q_{6x}^{\beta1}$ & $\frac{3 \sqrt{22} S[1]}{16} + \frac{\sqrt{55} S[3]}{16} + \frac{\sqrt{3} S[5]}{16}$ \\ 
 & $Q_{6y}^{\beta1}$ & $\frac{3 \sqrt{22} C[1]}{16} - \frac{\sqrt{55} C[3]}{16} + \frac{\sqrt{3} C[5]}{16}$ \\ 
 & $Q_{6z}^{\beta1}$ & $S[6]$ \\ \hline
$T_{2g}$ & $Q_{6x}^{\beta2}$ & $\frac{\sqrt{10} S[1]}{16} - \frac{9 S[3]}{16} + \frac{\sqrt{165} S[5]}{16}$ \\ 
 & $Q_{6y}^{\beta2}$ & $\frac{\sqrt{10} C[1]}{16} + \frac{9 C[3]}{16} + \frac{\sqrt{165} C[5]}{16}$ \\ 
 & $Q_{6z}^{\beta2}$ & $S[2]$ \\ 
\bottomrule
\end{tabular}
\hspace{1cm}\vspace{5mm}
\begin{tabular}{ccc}
\textbf{rank 7}\\
\toprule
irrep. & symbol & definition \\
\midrule
$A_{2u}$ & $Q_{7}^{\beta}$ & $\frac{\sqrt{78} S[2]}{12} + \frac{\sqrt{66} S[6]}{12}$ \\ \hline
$E_{u}$ & $Q_{7u}^{}$ & $S[4]$ \\ 
 & $Q_{7v}^{}$ & $\frac{\sqrt{66} S[2]}{12} - \frac{\sqrt{78} S[6]}{12}$ \\ \hline
$T_{1u}$ & $Q_{7x}^{\alpha1}$ & $- \frac{5 \sqrt{7} C[1]}{32} + \frac{3 \sqrt{21} C[3]}{32} - \frac{\sqrt{231} C[5]}{32} + \frac{\sqrt{429} C[7]}{32}$ \\ 
 & $Q_{7y}^{\alpha1}$ & $- \frac{5 \sqrt{7} S[1]}{32} - \frac{3 \sqrt{21} S[3]}{32} - \frac{\sqrt{231} S[5]}{32} - \frac{\sqrt{429} S[7]}{32}$ \\ 
 & $Q_{7z}^{\alpha1}$ & $C[0]$ \\ \hline
$T_{1u}$ & $Q_{7x}^{\alpha2}$ & $- \frac{3 \sqrt{33} C[1]}{32} - \frac{\sqrt{11} C[3]}{32} + \frac{25 C[5]}{32} + \frac{\sqrt{91} C[7]}{32}$ \\ 
 & $Q_{7y}^{\alpha2}$ & $- \frac{3 \sqrt{33} S[1]}{32} + \frac{\sqrt{11} S[3]}{32} + \frac{25 S[5]}{32} - \frac{\sqrt{91} S[7]}{32}$ \\ 
 & $Q_{7z}^{\alpha2}$ & $C[4]$ \\ \hline
$T_{2u}$ & $Q_{7x}^{\beta1}$ & $- \frac{\sqrt{858} C[1]}{64} - \frac{3 \sqrt{286} C[3]}{64} - \frac{5 \sqrt{26} C[5]}{64} - \frac{\sqrt{14} C[7]}{64}$ \\ 
 & $Q_{7y}^{\beta1}$ & $\frac{\sqrt{858} S[1]}{64} - \frac{3 \sqrt{286} S[3]}{64} + \frac{5 \sqrt{26} S[5]}{64} - \frac{\sqrt{14} S[7]}{64}$ \\ 
 & $Q_{7z}^{\beta1}$ & $C[6]$ \\ \hline
$T_{2u}$ & $Q_{7x}^{\beta2}$ & $- \frac{15 \sqrt{6} C[1]}{64} + \frac{19 \sqrt{2} C[3]}{64} - \frac{\sqrt{22} C[5]}{64} - \frac{\sqrt{2002} C[7]}{64}$ \\ 
& $Q_{7y}^{\beta2}$ & $\frac{15 \sqrt{6} S[1]}{64} + \frac{19 \sqrt{2} S[3]}{64} + \frac{\sqrt{22} S[5]}{64} - \frac{\sqrt{2002} S[7]}{64}$ \\ 
 & $Q_{7z}^{\beta2}$ & $C[2]$ \\ 
\bottomrule
\end{tabular}
\hspace{1cm}\vspace{5mm}
\begin{tabular}{ccc}
\textbf{rank 8}\\
\toprule
irrep. & symbol & definition \\
\midrule
$A_{1g}$ & $Q_{8}^{}$ & $\frac{\sqrt{33} C[0]}{8} + \frac{\sqrt{21} C[4]}{12} + \frac{\sqrt{195} C[8]}{24}$ \\ \hline
$E_{g}$ & $Q_{8u}^{1}$ & $- \frac{\sqrt{286} C[0]}{32} + \frac{\sqrt{182} C[4]}{16} + \frac{\sqrt{10} C[8]}{32}$ \\ 
 & $Q_{8v}^{1}$ & $C[6]$ \\ \hline
$E_{g}$ & $Q_{8u}^{2}$ & $- \frac{\sqrt{210} C[0]}{32} - \frac{\sqrt{330} C[4]}{48} + \frac{\sqrt{6006} C[8]}{96}$ \\ 
 & $Q_{8v}^{2}$ & $C[2]$ \\ \hline
$T_{1g}$ & $Q_{8x}^{\alpha1}$ & $- \frac{\sqrt{715} S[1]}{32} - \frac{\sqrt{273} S[3]}{32} - \frac{\sqrt{35} S[5]}{32} - \frac{S[7]}{32}$ \\ 
& $Q_{8y}^{\alpha1}$ & $\frac{\sqrt{715} C[1]}{32} - \frac{\sqrt{273} C[3]}{32} + \frac{\sqrt{35} C[5]}{32} - \frac{C[7]}{32}$ \\ 
 & $Q_{8z}^{\alpha1}$ & $S[8]$ \\ \hline
$T_{1g}$ & $Q_{8x}^{\alpha2}$ & $- \frac{\sqrt{77} S[1]}{32} + \frac{5 \sqrt{15} S[3]}{32} - \frac{3 \sqrt{13} S[5]}{32} - \frac{\sqrt{455} S[7]}{32}$ \\ 
& $Q_{8y}^{\alpha2}$ & $\frac{\sqrt{77} C[1]}{32} + \frac{5 \sqrt{15} C[3]}{32} + \frac{3 \sqrt{13} C[5]}{32} - \frac{\sqrt{455} C[7]}{32}$ \\ 
 & $Q_{8z}^{\alpha2}$ & $S[4]$ \\ \hline
$T_{2g}$ & $Q_{8x}^{\beta1}$ & $- \frac{\sqrt{858} S[1]}{64} + \frac{\sqrt{910} S[3]}{64} + \frac{7 \sqrt{42} S[5]}{64} + \frac{3 \sqrt{30} S[7]}{64}$ \\ 
& $Q_{8y}^{\beta1}$ & $- \frac{\sqrt{858} C[1]}{64} - \frac{\sqrt{910} C[3]}{64} + \frac{7 \sqrt{42} C[5]}{64} - \frac{3 \sqrt{30} C[7]}{64}$ \\ 
& $Q_{8z}^{\beta1}$ & $S[6]$ \\ \hline
$T_{2g}$ & $Q_{8x}^{\beta2}$ & $- \frac{\sqrt{70} S[1]}{64} + \frac{3 \sqrt{66} S[3]}{64} - \frac{\sqrt{1430} S[5]}{64} + \frac{\sqrt{2002} S[7]}{64}$ \\ 
& $Q_{8y}^{\beta2}$ & $- \frac{\sqrt{70} C[1]}{64} - \frac{3 \sqrt{66} C[3]}{64} - \frac{\sqrt{1430} C[5]}{64} - \frac{\sqrt{2002} C[7]}{64}$ \\ 
 & $Q_{8z}^{\beta2}$ & $S[2]$ \\ \bottomrule
\end{tabular}
\hspace{1cm}\vspace{5mm}
\begin{tabular}{ccc}
\textbf{rank 9}\\
\toprule
irrep. & symbol & definition \\
\midrule
$A_{1u}$ & $Q_{9}^{}$ & $\frac{\sqrt{102} S[4]}{12} - \frac{\sqrt{42} S[8]}{12}$ \\ \hline
$A_{2u}$ & $Q_{9}^{\beta}$ & $\frac{\sqrt{3} S[2]}{4} - \frac{\sqrt{13} S[6]}{4}$ \\ \hline
$E_{u}$ & $Q_{9u}^{}$ & $\frac{\sqrt{42} S[4]}{12} + \frac{\sqrt{102} S[8]}{12}$ \\ 
 & $Q_{9v}^{}$ & $-\frac{\sqrt{13} S[2]}{4} -\frac{\sqrt{3} S[6]}{4}$ \\ \hline
$T_{1u}$ & $Q_{9x}^{\alpha1}$ & $\frac{21 \sqrt{5} C[1]}{128} - \frac{\sqrt{2310} C[3]}{128} + \frac{3 \sqrt{286} C[5]}{128} - \frac{3 \sqrt{1430} C[7]}{256} + \frac{\sqrt{24310} C[9]}{256}$ \\ 
& $Q_{9y}^{\alpha1}$ & $\frac{21 \sqrt{5} S[1]}{128} + \frac{\sqrt{2310} S[3]}{128} + \frac{3 \sqrt{286} S[5]}{128} + \frac{3 \sqrt{1430} S[7]}{256} + \frac{\sqrt{24310} S[9]}{256}$ \\
\bottomrule
\end{tabular}
\hspace{1cm}\vspace{5mm}
\begin{tabular}{ccc}
\textbf{rank 10}\\
\toprule
irrep. & symbol & definition \\
\midrule
$A_{1g}$ & $Q_{10}^{}$ & $\frac{\sqrt{390} C[0]}{48} - \frac{\sqrt{22} C[4]}{8} - \frac{\sqrt{1122} C[8]}{48}$ \\ \hline
$A_{2g}$ & $Q_{10}^{\beta}$ & $- \frac{\sqrt{85} C[10]}{16} + \frac{\sqrt{1482} C[2]}{48} + \frac{\sqrt{57} C[6]}{48}$ \\ \hline
$E_{g}$ & $Q_{10u}^{1}$ & $\frac{11 \sqrt{420189} C[0]}{8988} + \frac{\sqrt{827645} C[4]}{1498} - \frac{\sqrt{146055} C[8]}{8988}$ \\ 
 & $Q_{10v}^{1}$ & $\frac{\sqrt{370006} C[10]}{749} + \frac{\sqrt{190995} C[2]}{749}$ \\ \hline
$E_{g}$ & $Q_{10u}^{2}$ & $\frac{3 \sqrt{3213210} C[0]}{11984} - \frac{83 \sqrt{1498} C[4]}{5992} + \frac{31 \sqrt{76398} C[8]}{11984}$ \\ 
& $Q_{10v}^{2}$ & $\frac{\sqrt{1209635} C[10]}{11984} - \frac{19 \sqrt{58422} C[2]}{35952} + \frac{\sqrt{2247} C[6]}{48}$ \\
\bottomrule
\end{tabular}
\hspace{1cm}\vspace{5mm}
\begin{tabular}{ccc}
\textbf{rank 11}\\
\toprule
irrep. & symbol & definition \\
\midrule
$A_{2u}$ & $Q_{11}^{\beta}$ & $\frac{\sqrt{798} S[10]}{48} + \frac{\sqrt{255} S[2]}{24} + \frac{3 \sqrt{6} S[6]}{16}$ \\ \hline
$E_{u}$ & $Q_{11u}^{1}$ & $S[8]$ \\ 
& $Q_{11v}^{1}$ & $- \frac{\sqrt{210} S[10]}{96} + \frac{\sqrt{969} S[2]}{48} - \frac{\sqrt{570} S[6]}{32}$ \\ \hline
$E_{u}$ & $Q_{11u}^{2}$ & $S[4]$ \\ 
 & $Q_{11v}^{2}$ & $- \frac{\sqrt{646} S[10]}{32} + \frac{\sqrt{35} S[2]}{16} + \frac{\sqrt{238} S[6]}{32}$ \\ \hline
$T_{1u}$ & $Q_{11x}^{\alpha1}$ & $\frac{\sqrt{88179} C[11]}{512} - \frac{21 \sqrt{66} C[1]}{512} + \frac{\sqrt{30030} C[3]}{512} - \frac{15 \sqrt{143} C[5]}{512} + \frac{\sqrt{36465} C[7]}{512} - \frac{\sqrt{46189} C[9]}{512}$ \\
\bottomrule
\end{tabular}

\end{center}

\subsection{Hexagonal multipoles up to rank 11}

\begin{center}
\renewcommand{\arraystretch}{1.1}
\begin{tabular}{ccc}
\textbf{rank 0}\\
\toprule
irrep. &      symbol & definition \\
\midrule
$A_{1g}$ &  $Q_{0}^{}$ &     $C[0]$ \\
\bottomrule
\end{tabular}
\hspace{1cm}\vspace{5mm}
\begin{tabular}{ccc}
\textbf{rank 1}\\
\toprule
irrep. &      symbol & definition \\
\midrule
$A_{2u}$ &  $Q_{z}^{}$ &     $C[0]$ \\ \hline
$E_{1u}$ &  $Q_{x}^{}$ &     $C[1]$ \\
&  $Q_{y}^{}$ &     $S[1]$ \\
\bottomrule
\end{tabular}
\hspace{1cm}\vspace{5mm}
\begin{tabular}{ccc}
\textbf{rank 2}\\
\toprule
irrep. &       symbol & definition \\
\midrule
$A_{1g}$ & $Q_{u}^{}$ & $C[0]$ \\ \hline
$E_{1g}$ & $Q_{zx}^{}$ & $C[1]$ \\ 
 & $Q_{yz}^{}$ & $S[1]$ \\ \hline
$E_{2g}$ & $Q_{v}^{}$ & $C[2]$ \\ 
 & $Q_{xy}^{}$ & $-S[2]$ \\ 
\bottomrule
\end{tabular}
\hspace{1cm}\vspace{5mm}
\begin{tabular}{ccc}
\textbf{rank 3}\\
\toprule
irrep. &            symbol & definition \\
\midrule
$A_{2u}$ & $Q_{3}^{\beta}$ & $C[0]$ \\ \hline
$B_{1u}$ & $Q_{3}^{\gamma}$ & $S[3]$ \\ \hline
$B_{2u}$ & $Q_{3}^{\delta}$ & $C[3]$ \\ \hline
$E_{1u}$ & $Q_{3u}^{\alpha}$ & $C[1]$ \\ 
 & $Q_{3v}^{\alpha}$ & $S[1]$ \\ \hline
$E_{2u}$ & $Q_{3u}^{\beta}$ & $C[2]$ \\ 
 & $Q_{3v}^{\beta}$ & $-S[2]$ \\ 
\bottomrule
\end{tabular}
\hspace{1cm}\vspace{5mm}
\begin{tabular}{ccc}
\textbf{rank 4}\\
\toprule
irrep. &             symbol & definition \\
\midrule
$A_{1g}$ & $Q_{4}^{}$ & $C[0]$ \\ \hline
$B_{1g}$ & $Q_{4}^{\gamma}$ & $C[3]$ \\ \hline
$B_{2g}$ & $Q_{4}^{\delta}$ & $S[3]$ \\ \hline
$E_{1g}$ & $Q_{4u}^{\alpha}$ & $C[1]$ \\ 
 & $Q_{4v}^{\alpha}$ & $S[1]$ \\ \hline
$E_{2g}$ & $Q_{4u}^{\beta1}$ & $C[4]$ \\ 
 & $Q_{4v}^{\beta1}$ & $S[4]$ \\ \hline
$E_{2g}$ & $Q_{4u}^{\beta2}$ & $C[2]$ \\ 
 & $Q_{4v}^{\beta2}$ & $-S[2]$ \\ 
\bottomrule
\end{tabular}
\hspace{1cm}\vspace{5mm}
\begin{tabular}{ccc}
\textbf{rank 5}\\
\toprule
irrep. &              symbol & definition \\
\midrule
$A_{2u}$ & $Q_{5}^{\beta}$ & $C[0]$ \\ \hline
$B_{1u}$ & $Q_{5}^{\gamma}$ & $S[3]$ \\ \hline
$B_{2u}$ & $Q_{5}^{\delta}$ & $C[3]$ \\ \hline
$E_{1u}$ & $Q_{5u}^{\alpha1}$ & $C[5]$ \\ 
 & $Q_{5v}^{\alpha1}$ & $-S[5]$ \\ \hline
$E_{1u}$ & $Q_{5u}^{\alpha2}$ & $C[1]$ \\ 
 & $Q_{5v}^{\alpha2}$ & $S[1]$ \\ \hline
$E_{2u}$ & $Q_{5u}^{\beta1}$ & $C[4]$ \\ 
 & $Q_{5v}^{\beta1}$ & $S[4]$ \\ \hline
$E_{2u}$ & $Q_{5u}^{\beta2}$ & $C[2]$ \\ 
 & $Q_{5v}^{\beta2}$ & $-S[2]$ \\ 
\bottomrule
\end{tabular}
\hspace{1cm}\vspace{5mm}
\begin{tabular}{ccc}
\textbf{rank 6}\\
\toprule
irrep. &              symbol & definition \\
\midrule
$A_{1g}$ & $Q_{6}^{1}$ & $C[0]$ \\ \hline
$A_{1g}$ & $Q_{6}^{2}$ & $C[6]$ \\ \hline
$A_{2g}$ & $Q_{6}^{\beta}$ & $S[6]$ \\ \hline
$B_{1g}$ & $Q_{6}^{\gamma}$ & $C[3]$ \\ \hline
$B_{2g}$ & $Q_{6}^{\delta}$ & $S[3]$ \\ \hline
$E_{1g}$ & $Q_{6u}^{\alpha1}$ & $C[5]$ \\ 
 & $Q_{6v}^{\alpha1}$ & $-S[5]$ \\ \hline
$E_{1g}$ & $Q_{6u}^{\alpha2}$ & $C[1]$ \\ 
 & $Q_{6v}^{\alpha2}$ & $S[1]$ \\ \hline
$E_{2g}$ & $Q_{6u}^{\beta1}$ & $C[4]$ \\ 
 & $Q_{6v}^{\beta1}$ & $S[4]$ \\ \hline
$E_{2g}$ & $Q_{6u}^{\beta2}$ & $C[2]$ \\ 
 & $Q_{6v}^{\beta2}$ & $-S[2]$ \\ 
\bottomrule
\end{tabular}
\hspace{1cm}\vspace{5mm}
\begin{tabular}{ccc}
\textbf{rank 7}\\
\toprule
irrep. &              symbol & definition \\
\midrule
$A_{1u}$ & $Q_{7}^{}$ & $S[6]$ \\ \hline
$A_{2u}$ & $Q_{7}^{\beta1}$ & $C[0]$ \\ \hline
$A_{2u}$ & $Q_{7}^{\beta2}$ & $C[6]$ \\ \hline
$B_{1u}$ & $Q_{7}^{\gamma}$ & $S[3]$ \\ \hline
$B_{2u}$ & $Q_{7}^{\delta}$ & $C[3]$ \\ \hline
$E_{1u}$ & $Q_{7u}^{\alpha1}$ & $C[7]$ \\ 
 & $Q_{7v}^{\alpha1}$ & $S[7]$ \\ \hline
$E_{1u}$ & $Q_{7u}^{\alpha2}$ & $C[5]$ \\ 
 & $Q_{7v}^{\alpha2}$ & $-S[5]$ \\ \hline
$E_{1u}$ & $Q_{7u}^{\alpha3}$ & $C[1]$ \\ 
 & $Q_{7v}^{\alpha3}$ & $S[1]$ \\ \hline
$E_{2u}$ & $Q_{7u}^{\beta1}$ & $C[4]$ \\ 
 & $Q_{7v}^{\beta1}$ & $S[4]$ \\ \hline
$E_{2u}$ & $Q_{7u}^{\beta2}$ & $C[2]$ \\ 
 & $Q_{7v}^{\beta2}$ & $-S[2]$ \\ 
\bottomrule
\end{tabular}
\hspace{1cm}\vspace{5mm}
\begin{tabular}{ccc}
\textbf{rank 8}\\
\toprule
irrep. &              symbol & definition \\
\midrule
$A_{1g}$ & $Q_{8}^{1}$ & $C[0]$ \\ \hline
$A_{1g}$ & $Q_{8}^{2}$ & $C[6]$ \\ \hline
$A_{2g}$ & $Q_{8}^{\beta}$ & $S[6]$ \\ \hline
$B_{1g}$ & $Q_{8}^{\gamma}$ & $C[3]$ \\ \hline
$B_{2g}$ & $Q_{8}^{\delta}$ & $S[3]$ \\ \hline
$E_{1g}$ & $Q_{8u}^{\alpha1}$ & $C[7]$ \\ 
 & $Q_{8v}^{\alpha1}$ & $S[7]$ \\ \hline
$E_{1g}$ & $Q_{8u}^{\alpha2}$ & $C[5]$ \\ 
 & $Q_{8v}^{\alpha2}$ & $-S[5]$ \\ \hline
$E_{1g}$ & $Q_{8u}^{\alpha3}$ & $C[1]$ \\ 
 & $Q_{8v}^{\alpha3}$ & $S[1]$ \\ \hline
$E_{2g}$ & $Q_{8u}^{\beta1}$ & $C[8]$ \\ 
 & $Q_{8v}^{\beta1}$ & $-S[8]$ \\ \hline
$E_{2g}$ & $Q_{8u}^{\beta2}$ & $C[4]$ \\ 
 & $Q_{8v}^{\beta2}$ & $S[4]$ \\ \hline
$E_{2g}$ & $Q_{8u}^{\beta3}$ & $C[2]$ \\ 
 & $Q_{8v}^{\beta3}$ & $-S[2]$ \\ 
\bottomrule
\end{tabular}
\hspace{1cm}\vspace{5mm}
\begin{tabular}{ccc}
\textbf{rank 9}\\
\toprule
irrep. &              symbol & definition \\
\midrule
$A_{1u}$ & $Q_{9}^{}$ & $S[6]$ \\ \hline
$A_{2u}$ & $Q_{9}^{\beta1}$ & $C[0]$ \\ \hline
$A_{2u}$ & $Q_{9}^{\beta2}$ & $C[6]$ \\ \hline
$B_{1u}$ & $Q_{9}^{\gamma1}$ & $S[9]$ \\ \hline
$B_{1u}$ & $Q_{9}^{\gamma2}$ & $S[3]$ \\ \hline
$B_{2u}$ & $Q_{9}^{\delta1}$ & $C[9]$ \\ \hline
$B_{2u}$ & $Q_{9}^{\delta2}$ & $C[3]$ \\ \hline
$E_{1u}$ & $Q_{9u}^{\alpha1}$ & $C[7]$ \\ 
 & $Q_{9v}^{\alpha1}$ & $S[7]$ \\ \hline
$E_{1u}$ & $Q_{9u}^{\alpha2}$ & $C[5]$ \\ 
 & $Q_{9v}^{\alpha2}$ & $-S[5]$ \\ \hline
$E_{1u}$ & $Q_{9u}^{\alpha3}$ & $C[1]$ \\ 
 & $Q_{9v}^{\alpha3}$ & $S[1]$ \\ \hline
$E_{2u}$ & $Q_{9u}^{\beta1}$ & $C[8]$ \\ 
 & $Q_{9v}^{\beta1}$ & $-S[8]$ \\ \hline
$E_{2u}$ & $Q_{9u}^{\beta2}$ & $C[4]$ \\ 
 & $Q_{9v}^{\beta2}$ & $S[4]$ \\ \hline
$E_{2u}$ & $Q_{9u}^{\beta3}$ & $C[2]$ \\ 
 & $Q_{9v}^{\beta3}$ & $-S[2]$ \\ 
\bottomrule
\end{tabular}
\hspace{1cm}\vspace{5mm}
\begin{tabular}{ccc}
\textbf{rank 10}\\
\toprule
irrep. &               symbol & definition \\
\midrule
$A_{1g}$ & $Q_{10}^{1}$ & $C[0]$ \\ \hline
$A_{1g}$ & $Q_{10}^{2}$ & $C[6]$ \\ \hline
$A_{2g}$ & $Q_{10}^{\beta}$ & $S[6]$ \\ \hline
$B_{1g}$ & $Q_{10}^{\gamma1}$ & $C[9]$ \\ \hline
$B_{1g}$ & $Q_{10}^{\gamma2}$ & $C[3]$ \\ \hline
$B_{2g}$ & $Q_{10}^{\delta1}$ & $S[9]$ \\ \hline
$B_{2g}$ & $Q_{10}^{\delta2}$ & $S[3]$ \\ \hline
$E_{1g}$ & $Q_{10u}^{\alpha1}$ & $C[7]$ \\ 
 & $Q_{10v}^{\alpha1}$ & $S[7]$ \\ \hline
$E_{1g}$ & $Q_{10u}^{\alpha2}$ & $C[5]$ \\ 
 & $Q_{10v}^{\alpha2}$ & $-S[5]$ \\ \hline
$E_{1g}$ & $Q_{10u}^{\alpha3}$ & $C[1]$ \\ 
 & $Q_{10v}^{\alpha3}$ & $S[1]$ \\ \hline
$E_{2g}$ & $Q_{10u}^{\beta1}$ & $C[10]$ \\ 
 & $Q_{10v}^{\beta1}$ & $S[10]$ \\ \hline
$E_{2g}$ & $Q_{10u}^{\beta2}$ & $C[8]$ \\ 
 & $Q_{10v}^{\beta2}$ & $-S[8]$ \\ \hline
$E_{2g}$ & $Q_{10u}^{\beta3}$ & $C[4]$ \\ 
 & $Q_{10v}^{\beta3}$ & $S[4]$ \\ \hline
$E_{2g}$ & $Q_{10u}^{\beta4}$ & $C[2]$ \\ 
 & $Q_{10v}^{\beta4}$ & $-S[2]$ \\ 
\bottomrule
\end{tabular}
\hspace{1cm}\vspace{5mm}
\begin{tabular}{ccc}
\textbf{rank 11}\\
\toprule
irrep. &               symbol & definition \\
\midrule
$A_{1u}$ & $Q_{11}^{}$ & $S[6]$ \\ \hline
$A_{2u}$ & $Q_{11}^{\beta1}$ & $C[0]$ \\ \hline
$A_{2u}$ & $Q_{11}^{\beta2}$ & $C[6]$ \\ \hline
$B_{1u}$ & $Q_{11}^{\gamma1}$ & $S[9]$ \\ \hline
$B_{1u}$ & $Q_{11}^{\gamma2}$ & $S[3]$ \\ \hline
$B_{2u}$ & $Q_{11}^{\delta1}$ & $C[9]$ \\ \hline
$B_{2u}$ & $Q_{11}^{\delta2}$ & $C[3]$ \\ \hline
$E_{1u}$ & $Q_{11u}^{\alpha1}$ & $C[11]$ \\ 
 & $Q_{11v}^{\alpha1}$ & $-S[11]$ \\ \hline
$E_{1u}$ & $Q_{11u}^{\alpha2}$ & $C[7]$ \\ 
 & $Q_{11v}^{\alpha2}$ & $S[7]$ \\ \hline
$E_{1u}$ & $Q_{11u}^{\alpha3}$ & $C[5]$ \\ 
 & $Q_{11v}^{\alpha3}$ & $-S[5]$ \\ \hline
$E_{1u}$ & $Q_{11u}^{\alpha4}$ & $C[1]$ \\ 
 & $Q_{11v}^{\alpha4}$ & $S[1]$ \\ \hline
$E_{2u}$ & $Q_{11u}^{\beta1}$ & $C[10]$ \\ 
 & $Q_{11v}^{\beta1}$ & $S[10]$ \\ \hline
$E_{2u}$ & $Q_{11u}^{\beta2}$ & $C[8]$ \\ 
 & $Q_{11v}^{\beta2}$ & $-S[8]$ \\ \hline
$E_{2u}$ & $Q_{11u}^{\beta3}$ & $C[4]$ \\ 
 & $Q_{11v}^{\beta3}$ & $S[4]$ \\ \hline
$E_{2u}$ & $Q_{11u}^{\beta4}$ & $C[2]$ \\ 
 & $Q_{11v}^{\beta4}$ & $-S[2]$ \\ 
\bottomrule
\end{tabular}

\end{center}

\end{document}